\documentclass[journal,10pt]{IEEEtran}

\usepackage{silence}

\usepackage{premable_doc}
\usepackage{premable_tikz}
\usepackage{fancyhdr}  

\newacronym{npd}{NPD}{neural polar decoder}
\newacronym{hy}{HY}{Honda-Yamamoto}
\newacronym{bms}{BMS}{binary memoryless symmetric}
\newcommand{\nnc}[3]{\cG_\mathsf{NN}^{\left(#1, #2, #3\right)}}

\begin{document}

\title{A Study of Neural Polar Decoders for Communication}

\author{Rom Hirsch,~\IEEEmembership{Student Member,~IEEE,}
        Ziv Aharoni,~\IEEEmembership{Member,~IEEE,}
        Henry D. Pfister,~\IEEEmembership{Senior Member,~IEEE,}
        and Haim H. Permuter,~\IEEEmembership{Senior Member,~IEEE}%
\thanks{Rom Hirsch and Haim H. Permuter are with the Department of Electrical and Computer Engineering, Ben-Gurion University of the Negev, Beer-Sheva, Israel.}%
\thanks{Ziv Aharoni and Henry D. Pfister are with the Department of Electrical and Computer Engineering, Duke University, Durham, NC, USA.}
}

\maketitle

\begin{abstract}

In this paper, we adapt and analyze Neural Polar Decoders (NPDs) for end-to-end communication systems. While prior work demonstrated the effectiveness of NPDs on synthetic channels, this study extends the NPD to real-world communication systems. The NPD was adapted to complete OFDM and single-carrier communication systems. To satisfy practical system requirements, the NPD is extended to support any code length via rate matching, higher-order modulations, and robustness across diverse channel conditions. The NPD operates directly on channels with memory, exploiting their structure to achieve higher data rates without requiring pilots and a cyclic prefix. Although NPD entails higher computational complexity than the standard 5G polar decoder, its neural network architecture enables an efficient representation of channel statistics, resulting in manageable complexity suitable for practical systems. Experimental results over 5G channels demonstrate that the NPD consistently outperforms the 5G polar decoder in terms of BER, BLER, and throughput. These improvements are particularly significant for low-rate and short-block configurations, which are prevalent in 5G control channels. Furthermore, NPDs applied to single-carrier systems offer performance comparable to OFDM with lower PAPR, enabling effective single-carrier transmission over 5G channels. These results position the NPD as a high-performance, pilotless, and robust decoding solution.

\end{abstract}


\glsresetall
\blfootnote{This research was supported in part by a donation from Nokia Solutions and Networks, in part by the National Science Foundation (NSF) under Grant 2308445, and in part by the NSF–Israel Binational Science Foundation (BSF) under Grant 3211/23. Any opinions, findings, recommendations, and conclusions are those of the authors and do not necessarily reflect the views of these sponsors. Source code 
is available at: \url{https://github.com/romhirsch/npd-communication}.}

\section{Introduction}

\par Polar codes, introduced by Arıkan in 2009 \cite{arikan2009channel}, are the first class of codes proven to achieve the capacity of symmetric binary-input discrete memoryless channels (B-DMCs) under low-complexity successive cancellation (SC) decoding. In 5G, polar codes are primarily used for control channels, where high performance is required with a low rate and short code length. Their inclusion in the 5G New Radio (NR) standard for uplink and downlink control information, use cases such as enhanced mobile broadband (eMBB) and broadcast channel (BCH) highlight their practical relevance in modern wireless communication systems. 

\par In wireless communication, memory effects are prevalent in the form of intersymbol interference (ISI) from multipath propagation and time variations induced by user mobility. Current 5G systems favor low-complexity decoders designed for memoryless channels, which limits their performance in wireless channels. To enable the use of such decoders in wireless channels, traditional solutions, such as orthogonal frequency-division multiplexing (OFDM), interleaving, and equalization, aim to transform the channel into a memoryless channel. However, these methods introduce additional system complexity, such as channel estimation and equalization, as well as overhead from pilot symbols and the cyclic prefix (CP). Consequently, 5G systems achieve reliability at the expense of spectral efficiency and data rate.

\par The literature presents numerous studies that address this challenge using neural network-based communication systems. These systems aim to transform the system into a memoryless system for coding techniques. Notable examples include \cite{aoudia2022end, doerner2017deep, huleihel2024low, aoudia2022waveform, ye2020deep, felix2018ofdm}, which aim to improve reliability while minimizing pilot overhead. These studies proposed models capable of estimating log-likelihood ratios (LLRs) over wireless channels while minimizing or eliminating the use of pilots. However, they rely on end-to-end architectures, which present practical challenges, such as the need for a differentiable channel model during training.

\par In contrast, this work aims to tackle the problem by applying coding techniques directly over channels with memory, specifically for 5G polar code scenarios. Prior studies have extended polar coding to memory channels, such as the successive cancellation trellis (SCT) decoder \cite{wang2015construction}, which achieves optimal decoding performance for finite-state channels (FSCs). However, the SCT decoder is not practically applicable in 5G scenarios because of its computational complexity, which scales cubically with the channel state size, expressed as $O(|S|^3 N \log N)$, where $|S|$ denotes the number of channel states. For instance, in a memory channel with 23 taps and BPSK modulation, $|S| = 2^{23}$, which leads to computationally infeasible SCT decoding. Moreover, the SCT decoder assumes full knowledge of the channel model, which is significantly more difficult to estimate in memory channels than in memoryless channels.

Recently, to address SCT limitations, NPDs have been proposed \cite{aharoni2023data_arcive,aharoniCodeRateOptimization2024}, enabling direct coding over memory channels with high reliability. NPDs retain the recursive structure of successive cancellation (SC) decoding but replace the core operations with neural networks (NNs). Unlike the SCT decoder, the NPDs computational complexity $O(dhN \log N)$ is independent of the channel memory size, where $d$ and $h$ are NN parameters (input and hidden dimensions). Furthermore, NPDs do not require prior knowledge of the channel model because they are trained on input-output observations. These properties make NPDs attractive for wireless channels.

\par  Despite these advances, several practical gaps remain. Existing NPDs do not fully support the essential features of communication systems, such as high-order modulation, rate matching, and generalization across diverse channel conditions and signal-to-noise ratios (SNRs). Furthermore, the NPD was benchmarked only on synthetic channels and did not present integration into end-to-end communication systems.

\par In this work, we address these gaps by extending the NPD to fully support modern communication system requirements. We developed a generalized NPD architecture that supports high-order modulations, rate-matching mechanisms, and robust performance across a wide range of SNRs and channel conditions. To demonstrate practical viability, the NPD was adapted to complete end-to-end communication systems based on both OFDM and single-carrier waveforms.

\par By evaluating the proposed NPD over the 5G channels using tapped-delay line (TDL) models, we demonstrate that it consistently outperforms the standardized 5G polar decoder in terms of bit error rate (BER), block error rate (BLER), and throughput even without pilots and CP. The performance gains are particularly notable at low code rates and short block lengths, which align well with the requirements of 5G control channels. This advantage persists under challenging conditions, including diverse delay profiles, large delay spreads, Doppler shifts, and nonlinear distortions. Furthermore, our results show that the NPD enables reliable single-carrier communication over 5G channels, while single-carrier waveforms offer advantages over OFDM, including lower peak-to-average power ratio (PAPR) and reduced hardware complexity. Although this enhanced performance comes at the cost of increased complexity relative to 5G polar codes and requires scenario-specific designs, it remains practical and well-suited for systems with greater computational resources and stringent requirements for high reliability.

\subsection{Contributions}

This study makes the following key contributions:

\begin{enumerate}
    \item We show how to adapt NPD into end-to-end communication systems and demonstrate their effectiveness with both OFDM and single-carrier waveforms.

    \item We develop rate-matching for NPD, enabling compatibility with any code lengths.

    \item We extend the NPD to support high-order modulation schemes. 

    \item We adapt the NPD architecture, training procedure, and design to achieve robustness over a wide range of SNRs and channel conditions. Remarkably, a single trained model generalizes effectively across diverse scenarios without requiring fine-tuning or retraining.
        
    \item Simulations over 5G channels demonstrate that the proposed NPD outperforms the 5G polar decoder in terms of BER, BLER, and throughput, even in the absence of pilots and CP. It offers significant gains for short block lengths and low code rates, making it well-suited for 5G control channels.
    
\end{enumerate}

\subsection{Organization}
The remainder of this paper is organized as follows. Section \ref{sec:pre} outlines the notations and preliminaries. Section \ref{sec:npd} provides a detailed description of NPDs for communication, highlighting the key contributions of this study, including rate matching for NPD, an adapted NPD structure for communication, introduction of end-to-end communication systems using NPD, and a modified training and design procedure. Section \ref{sec:exp} presents the experimental results over 5G channels and discusses the insights derived from them. Finally, Section \ref{sec:conc} concludes the paper by summarizing the main contributions and proposing directions for future research.

\section{Notations and Preliminaries} \label{sec:pre}
Throughout this paper, random variables (RVs) are denoted by capital letters, and their realizations are denoted by lower-case letters, for example, $X$ and $x$. Calligraphic letters denote sets, for example, $\mathcal{X}$. We use the notation $X_i^j$ to denote the RV $(X_i,X_{i+1},\dots,X_j)$ and $x_i^j$ to denote its realization. If $i=1$, we may omit the index $i$ to simplify the notation, that is, $X^j$. The probability $\text{Pr}[X=x]$ is denoted as $P_X(x)$. The Mutual information (MI) between two RVs $X,Y$ is denoted by $\sI(X;Y)$. Class of neural networks with input $d_i$ and output $d_0$ dimensions and $h\in \mathbb{N}$ neurons denoted by $\cG_\mathsf{NN}^{(d_i,h,d_o)}$.

\subsection{Communication Channel Model}\label{sec:communication:model}

This section describes the communication channel model used to evaluate the performance of the proposed NPD for communication. Specifically, we consider wireless communication channels modeled as time-varying multipath fading channels that induce fluctuations in the amplitude, phase, and angle of arrival of the received signal. These impairments are characteristic of practical wireless systems. The following model is based on the discrete-time baseband model for the wireless channel presented in \cite{fundamentalsWireless}.

\begin{definition}[Wireless channel model]\label{def:channelmodel}
Let $s:\mathbb{Z}\rightarrow\mathbb{C}$ denote the complex baseband signal, where \( s[t] \in \mathbb{C} \) is the symbol transmitted at the discrete time index \( t \). In OFDM systems, the transmitted signal is generated by mapping the encoded bits to modulation symbols, followed by OFDM modulation, which includes applying an inverse fast Fourier transform (IFFT) and appending a CP. In single-carrier systems, \( s[t] \) corresponds directly to the modulated symbol derived from the encoded bits. The received signal \( y[t] \in \mathbb{C} \) is modeled as the output of a time-varying multipath fading channel, expressed as
\begin{equation}
    y[t] = \sum_{\ell=L_{min}}^{L_{max}} s[t - \ell] \cdot h_\ell[t]  + w[t],
\end{equation}
where \( w[t] \sim \mathcal{CN}(0, \sigma^2) \) denotes the complex additive white Gaussian noise (AWGN), at time $t$. The parameters $L_{min}$ and $L_{max}$ are the smallest and largest time lags, respectively, for the discrete time channel model. The channel impulse response at time \( t \) is denoted by \( h_l[t] \) and modeled by
\begin{equation}
    h_\ell[t] = \sum_{p=0}^{P-1} a_p(t/W) \cdot \mathrm{sinc}(\ell - W \tau_p(t)),
\end{equation}
where \( P \) is the number of propagation paths (taps), \( a_p(t) \in \mathbb{C} \) is the complex gain of the \( p \)-th path at time \( t \), \( \tau_p(t) \in \mathbb{R} \) is the corresponding delay, and \( W \) is the system bandwidth. The sinc function is given by \( \mathrm{sinc}(x) = \frac{\sin(\pi x)}{\pi x} \). This model assumes ideal pulse shaping and matched filtering using a sinc pulse.

\end{definition}

To emulate realistic wireless propagation conditions, we adopted TDL channel models standardized by 3GPP \cite{3gpp38901}. These models define the delay profiles and power levels for each tap, capturing the effects of multipath propagation. There are five standardized TDL models: TDL-A, TDL-B, and TDL-C represent non-line-of-sight (NLOS) scenarios, whereas TDL-D and TDL-E are designed for line-of-sight (LOS) conditions.

\subsection{Polar Codes}

\subsubsection{Polar Transform} Let $G_N = B_N F^{\otimes n}$ represent the generator matrix for a code length of $N=2^n$, where $n \in \bN$, defining what is known as Ar\i{}kan's polar transform.
Matrix $B_N$ is the permutation matrix associated with the bit-reversal permutation. It is defined by the recursive relation $B_N = R_N(I_2 \otimes B_{N/2})$ starting from $B_2=I_2$. The term $I_N$ denotes the identity matrix of size $N$ and $R_N$ denotes a permutation matrix called reverse-shuffle \cite{arikan2009channel}. The matrix $G_N$ satisfies $G_NG_N=I_N$.
The term $A\otimes B$ denotes the Kronecker product of $A$ and $B$. The term $A^{\otimes n}:=A\otimes A\otimes\dots\otimes A$ denotes the application of the $\otimes$ operator $n$ times.

\subsubsection{Polar Code} 
We define a polar code by the tuple $\left(\cX, \cY, W, E, F, G, H\right)$ that contains the channel input alphabet $\cX$, channel output alphabet $\cY$, channel $W$, the channel embedding $E$ and the core components of the \gls{sc} decoder, $F,G,H$. The term $E:\cY\to\cE$ denotes the channel embedding, where $\cE\subset\bR^d$. The functions $F:\cE\times\cE\to\cE, \;G:\cE\times\cE\times\cX\to\cE$ denote the check-node and bit-node operations, respectively. We denote by $H:\cE\to\bR$ a mapping of the embedding into an LLR value, that is, a soft-decision.

\subsubsection{Polar Codes for Memoryless Channels}

Let \( W \) be a memoryless channel. The channel embedding \( E \) is defined as
\begin{equation}\label{eqn:memoryless-channel-stats}
    E(y) = \log\frac{\f{W}{y|1}}{\f{W}{y|0}} + \log\frac{\f{P_{X}}{1}}{\f{P_{X}}{0}},
\end{equation}
where the second term cancels out for a uniform input \( P_X \). The SC decoding operations are given by
\begin{align}\label{eqn:sc_ops}
    &F(e_1, e_2) =  -2\tanh^{-1}\left(\tanh{\frac{e_1}{2}}\tanh{\frac{e_2}{2}}\right), \nn\\
    &G(e_1, e_2, u) =  e_2 + (-1)^{u}e_1, \nn\\
    &\hspace{0.1cm}H(e_1) = e_1,
\end{align}
where \( e_1, e_2 \in \mathbb{R} \), \( u \in \mathcal{X} \). The hard decision function is \( h(l) = \mathbb{I}_{l > 0} \), where \( \mathbb{I} \) is the indicator function.

 SC decoding \cite{arikan2009channel} yields an estimate of transmitted bits, specifically, given the channel outputs \( y_1^N \), the SC decoding performs the map
\[
(y_1^N, f_1^N) \mapsto \left\{\hat{u}_i, \f{L_{U_i|U_1^{i-1},Y_1^N}}{\hat{u}_1^{i-1},y_1^N}\right\}_{i \in [N]},
\]
where the LLR of synthetic channels is defined as
\begin{equation}\label{eqn:llr_def}
    \f{L_{U_i|U_1^{i-1},Y_1^N}}{u_1^{i-1},y_1^N} = \log\frac{\f{P_{U_i|U_1^{i-1},Y_1^N}}{1| u_1^{i-1},y_1^N}}{\f{P_{U_i|U_1^{i-1},Y_1^N}}{0| u_1^{i-1},y_1^N}}.
\end{equation}

The vector \( f_1^N \) represents the frozen bits that are shared between the encoder and decoder:
\begin{equation}\label{eqn:fN}
f_i = 
\begin{cases}
    u_i & i \in \mathcal{F}, \\
    0.5 & i \in \mathcal{A},
\end{cases}
\end{equation}

where the value $0.5$ is used, by convention, to indicate that the bit needs to be decoded, the set $\cA\subseteq [N]$ is the information set, and $\cF=[N]\setminus\cA$ is the frozen set. 
For more details on \gls{sc} decoding, the reader may refer to \cite[Section VIII]{arikan2009channel}.

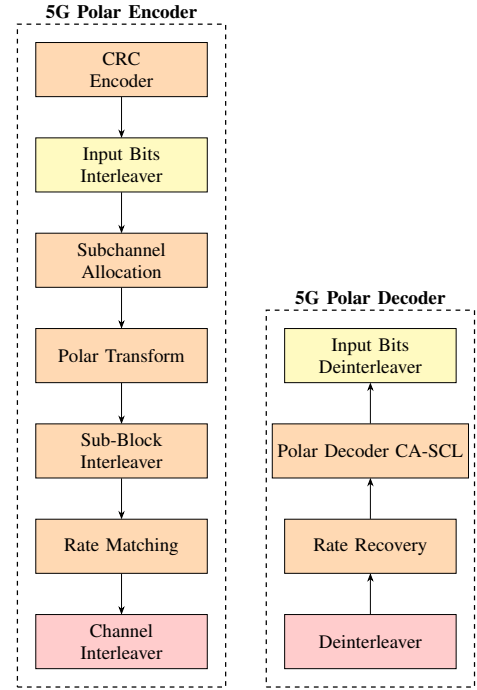
\begin{figure}[h!]
\centering
\scalebox{0.6}{
\begin{tikzpicture}[>=Stealth, node distance=0.9cm and 1.7cm, font=\large]

\tikzstyle{block} = [draw, rectangle, minimum width=3.8cm, minimum height=1.2cm, align=center]
\tikzstyle{rblock} = [block, fill=red!20]
\tikzstyle{oblock} = [block, fill=orange!30]
\tikzstyle{yblock} = [block, fill=yellow!30]

\node[oblock] (crc) {CRC\\Encoder};
\node[yblock, below=of crc] (interleaver) {Input Bits\\Interleaver};
\node[oblock, below=of interleaver] (alloc) {Subchannel\\Allocation};
\node[oblock, below=of alloc] (polar) {Polar Transform};
\node[oblock, below=of polar] (subint) {Sub-Block\\Interleaver};
\node[oblock, below=of subint] (rm) {Rate Matching};
\node[rblock, below=of rm] (chint) {Channel\\Interleaver};

\node[rblock, right=of chint] (deint) {Deinterleaver};
\node[oblock, above=of deint] (rr) {Rate Recovery};
\node[oblock, above=of rr] (decoder) {Polar Decoder CA-SCL};
\node[yblock, above=of decoder] (iildeint) {Input Bits\\Deinterleaver};

\draw[->] (crc) -- (interleaver);
\draw[->] (interleaver) -- (alloc);
\draw[->] (alloc) -- (polar);
\draw[->] (polar) -- (subint);
\draw[->] (subint) -- (rm);
\draw[->] (rm) -- (chint);

\draw[->] (deint) -- (rr);
\draw[->] (rr) -- (decoder);
\draw[->] (decoder) -- (iildeint);

\node[draw=black, thick, dashed, fit=(crc)(chint), inner sep=0.4cm, label=above:\textcolor{black}{\textbf{5G Polar Encoder}}] {};

\node[draw=black, thick, dashed, fit=(deint)(iildeint), inner sep=0.4cm, label=above:\textcolor{black}{\textbf{5G Polar Decoder}}] {};

\end{tikzpicture}
}
\caption{5G Polar code system. Yellow, red, and orange blocks are implemented in downlink, uplink, and both, respectively.}
\label{fig:5g_polar_2col}
\end{figure}

\subsubsection{5G Polar Code}

The 5G polar encoder begins with a cyclic redundancy check (CRC) encoder that appends a 24-bit (downlink) or 11-bit (uplink) CRC to the input bits. In the downlink, an input-bit interleaver is applied before subchannel allocation, which constructs the polar input vector \( u^N \) by placing the information and CRC bits in the information set \( \mathcal{A} \) and zeros in the frozen set \( \mathcal{F} \). The polar transform then generates the codeword \( x^N = u^N F^{\otimes n} \). Notably, the bit-reversal permutation \( B_N \) is not included. The codeword is then passed through a subblock interleaver and a rate-matching unit, which adjusts the length to the desired number of transmission bits \( N_r \leq N \) via puncturing, shortening, or repetition~\cite{richardson2008modern}. In the uplink, the bits are further permuted by a channel interleaver, whereas in the downlink, this step is omitted.

In the 5G polar decoder, processing begins with a deinterleaver in the uplink, whereas this step is omitted in the downlink. This is followed by a rate recovery stage that reconstructs the original codeword length \( N \), resulting in a length-\( N \) LLR vector. These LLRs are then decoded using CRC-aided successive cancellation list (CA-SCL) decoding~\cite{niu2012crc}. In the downlink, an additional input bit deinterleaver is applied after decoding to restore the original bit order. For further details, refer to~\cite{3gpp38212,bioglio2020design}.

\subsubsection{Puncturing for Polar Codes}

Puncturing~\cite{shuval2024strong,low-complex-puncturing,BeyondRate, DEGA} is a rate-matching technique used to reduce the length of a polar code from an original length $N$ to a target length $N_r \leq N$ by omitting the transmission of $P=N - N_r$ bits. At the decoder, these punctured bits are treated as erasures, and decoding is performed on the original length-$N$ codeword, with the punctured bits appropriately handled.

\begin{definition}[Puncturing Set Based on Bit-Reversal Permutation]\label{def:puncset}
The puncturing set $\mathcal{P}$ is defined as the set of indices corresponding to the first $P$ positions of the bit-reversed sequence $\mathcal{P}=B_N(1, \dots, P)$. Specifically, for example, with parameters $N=8$ and $P=3$, the bit-reversed sequence is $B_N=(1, 5, 3, 7, 2, 6, 4, 8)$, resulting in the puncturing set $\mathcal{P}=\{1,5,3\}$.
\end{definition}

For instance, in \cite{low-complex-puncturing}, punctured bits are determined by a puncturing set defined as $\mathcal{P} = B_N(1, \dots, P)$ as define in Definition \ref{def:puncset}. Consequently, at the encoder, the first $P$ bits of the codeword $x^N = u^N G_N$, arranged according to bit-reversal ordering, are punctured. At the receiver, LLRs are computed for the received symbols as usual, while the LLRs at the punctured positions are set to zero to reflect the prior of the bit $\log(\frac{P_X(1)}{P_X(0)})$.

\subsection{Neural Polar Decoder}
The NPD introduced by \cite{aharoni2023data_arcive} performs SC decoding by replacing the functions $E,F,G,H$ with NNs. In the definitions that follow, we separate the description of $E$ from the $F,G,H$ functions for purposes that will be detailed in Section~\ref{sec:npd:structure}. Because the NPD preserves the structure of the SC decoder, it can be naturally extended to successive cancellation list (SCL) and CRC-aided SCL (CA-SCL) decoding, as shown in~\cite{aharoniCodeRateOptimization2024, tal2015list}. The NPD with list-size $L$ has a computational complexity of \( O(L h d N \log N) \).

\begin{definition}[Channel Embedding]\label{def:embedding}
    Let $E_{\theta_E}:\cY\to \bR^d$ be the \gls{nn} that embeds the channel outputs into the embedding space of the \gls{npd}, where $d\in\bN$ is the dimension of the embedding space. For $\cY \subset \bR$, we define $E_{\theta_E} \in \nnc{1}{h}{d}$ as the embedding \gls{nn}, where $h\in\bN$ is the number of hidden neurons.
\end{definition}

\begin{definition}[NPD SC functions]\label{def:npd}
Let $F_{\theta_F}$, $G_{\theta_F}$, $H_{\theta_H}$ be the \glspl{nn} components of the \gls{npd}. The \gls{npd}'s \glspl{nn} are defined by 
\begin{itemize}
    \item $F_{\theta_F} \in \nnc{2d}{h}{d}$ is the check-node \gls{nn},
    \item $G_{\theta_G} \in \nnc{2d+1}{h}{d}$ is the bit-node \gls{nn},
    \item $H_{\theta_H} \in \nnc{d}{h}{1}$ is the embedding-to-LLR \gls{nn},
\end{itemize}
where $d\in\bN$ is the dimension of the embedding space and $h\in\bN$ is the number of hidden neurons.
\end{definition}

\section{Neural Polar Decoders for Communication}\label{sec:npd}
This section describes the NPD for communication systems, building upon the original NPD introduced in~\cite{aharoni2023data_arcive, aharoniCodeRateOptimization2024}. We present several key enhancements to previous studies. First, the NPD is adapted to support higher-order modulation schemes and improved robustness across a wide range of signal-to-noise ratios (SNRs). Second, compatibility with rate matching is achieved by integrating a puncturing mechanism into the NPD architecture. Third,  we outline the NPD-based communication system for both single-carrier and OFDM-based systems. Finally, we describe the adapted training and design procedures used to optimize the performance under practical communication scenarios.

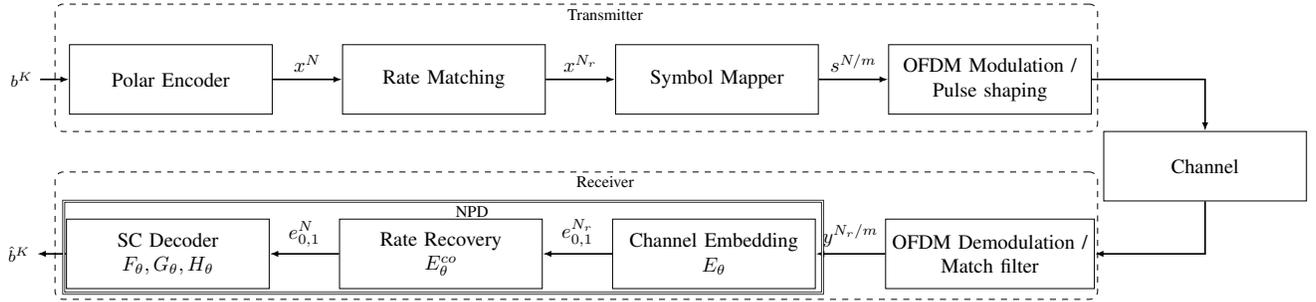
\begin{figure*}[h!]
    \centering
    \scalebox{0.77}{\tikzstyle{block} = [rectangle, draw, minimum width=3.5cm, minimum height=1.2cm, align=center, font=\normalsize]
\tikzstyle{arrow} = [-{Latex[length=1.5mm]}, thick]
\tikzstyle{sectionbox} = [draw, dashed, inner sep=10pt, rounded corners]

\begin{tikzpicture}[node distance=1.8cm and 1.2cm, every node/.style={transform shape}]

  \node[block] (polarEnc) {Polar Encoder};
  \node[block, right=of polarEnc] (rm) {Rate Matching};
  \node[block, right=of rm] (mapper) {Symbol Mapper};
  \node[block, right=of mapper] (ofdmMod) {OFDM Modulation / \\ Pulse shaping};

  \node[block, below=of ofdmMod] (ofdmDemod) {OFDM Demodulation  / \\ Match filter};
  \node[block, left=of ofdmDemod] (channel_emb) {Channel Embedding \\ $E_\theta$};
  \node[block, left=of channel_emb] (rm_np) {Rate Recovery \\ $E_\theta^{co}$};
  \node[block, left=of rm_np] (sc_npd) {SC Decoder \\ $F_\theta, G_\theta, H_\theta$};
  \node[left=of sc_npd, xshift=0.7cm](bl) {\footnotesize $\hat{b}^K$};
  \coordinate[left=of polarEnc, xshift=0.7cm] (input);
  
  \coordinate[right=of sc_npd] (output);

  \node[block, right=of ofdmMod,xshift=-1cm, yshift=-1.5cm] (channel) {Channel};

  \draw[arrow] (input) node[left] {\footnotesize $b^K$} -- (polarEnc);
  \draw[arrow] (polarEnc) -- (rm) node[midway,above] () {$x^N$};
  \draw[arrow] (rm) -- (mapper) node[midway,above] () {$x^{N_r}$};
  \draw[arrow] (mapper) -- (ofdmMod) node[midway,above] () {$s^{N/m}$};
  \draw[arrow] (ofdmMod) -| (channel) node[midway,above, xshift=-1cm] (){}; 
  \draw[arrow] (channel) |- (ofdmDemod) node[midway,above, xshift=-1cm](){}; 
  \draw[arrow] (ofdmDemod) -- (channel_emb) node[midway,above] () {$y^{N_r/m}$};
  \draw[arrow] (channel_emb) -- (rm_np) node[midway,above] () {$e_{0,1}^{N_r}$};
  \draw[arrow] (rm_np) -- (sc_npd) node[midway,above] () {$e_{0,1}^N$} ;
  \draw[arrow] (sc_npd.west) -- (bl) ;


    \node[sectionbox, minimum width=18cm, minimum height=2.2cm, label={[xshift=0.5cm, yshift=-0.4cm] \footnotesize Transmitter}] at (7, 0.2) {};

    \node[sectionbox, minimum width=18cm, minimum height=2.2cm, label={[xshift=0.5cm, yshift=-0.4cm] \footnotesize Receiver}] at (7, -2.7) {};
    
\node[draw, double, minimum width=13.1cm, minimum height=1.59cm, label={[xshift=0.5cm, yshift=-0.4cm] \footnotesize NPD}] at (4.7, -2.9) {};


\end{tikzpicture}}
    \caption{Block diagram of an end-to-end communication system integrating the proposed NPD. The diagram of the system presents both OFDM and single-carrier waveforms, which differ only in their Waveform block.}
    \label{fig:combined_systems}
\end{figure*} 

\subsection{Channel Embedding for Communication}\label{sec:npd:structure}

Building on the original NPD architecture, we modified the NN channel embedding to introduce two key enhancements that improve its suitability for practical communication systems.

First, to support higher-order modulations, the channel-embedding function is modified to produce multiple embedding vectors per received symbol. Specifically, it outputs one embedding per bit modulated in the symbol, allowing the decoder to operate effectively with modulation formats beyond BPSK (e.g., QPSK and 16-QAM).

Second, to improve the robustness across a wide range of SNRs, we incorporated the estimated noise variance \( N_0 \in \mathbb{R}^+ \) as an auxiliary input to the embedding network. This allows the decoder to adapt to varying channel conditions during inference, without retraining.

Importantly, the remaining neural functions used in decoding \( F_\theta \), \( G_\theta \), and \( H_\theta \) are shared across all SNR levels and modulation schemes. Definition~\ref{def:embedding_communication} formally specifies the extended embedding function used in the proposed NPD.

\begin{definition}[Channel Embedding for Communication]\label{def:embedding_communication}
Let \( E_{\theta_E}: \mathbb{R}^2 \times \mathbb{R}^+ \to \mathbb{R}^{d \cdot m} \) be a neural network that takes as input the real and imaginary parts of the received symbol \( y \in \mathbb{C} \), along with the estimated noise variance \( N_0 \in \mathbb{R}^+ \). The output is a vector in \( \mathbb{R}^{d \cdot m} \), which is reshaped into \( m \) embeddings of dimension \( d \), where \( m \) corresponds to the number of bits per symbol. The embedding network is parameterized as \( E_{\theta_E} \in \mathcal{G}^{(3, h, dm)}_{\text{NN}} \), where \( h \in \mathbb{N} \) denotes the number of neurons units.
\end{definition}

\subsection{Rate Matching for NPD}\label{sec:rm}

Polar codes were originally designed for code lengths $N$ that are powers of two. However, practical systems require flexibility to accommodate any code length. To address this, we adapted \emph{puncturing} strategies to the NPD for communication systems.

At the encoder for Binary Phase Shift Keying (BPSK) modulation, we adopt the punctured polar code from \cite{low-complex-puncturing}, removing $P$ bits according to the puncturing set $\mathcal{P}=B_N(1,\dots,P)$, as described in Definition \ref{def:puncset}. For higher-order modulation schemes, we puncture entire symbols corresponding to bits defined by the puncturing set $\mathcal{P}=B_N(1,\dots,P/m)$, where $m$ represents the number of bits per symbol.  For example, given $N=8$, $P=4$, and $m=2$ bits per symbol, the bit-reversal order is $B_N=(1,5,3,7,2,6,4,8)$. Thus, bits 1 and 5 determine the symbols to puncture, leading us to puncture the symbols containing bits $\{1,2\}$ and $\{4,5\}$. Consequently, the puncturing set becomes $\mathcal{P}=\{1,2,4,5\}$, indicating the bits that are not transmitted. This heuristic approach for higher-order modulation achieved consistently good performance in our experiments, with potential enhancements to be investigated in future work.

At the decoder, we enhance the NPD by introducing a learned \emph{constant embedding} to represent punctured bits. We define \emph{constant embeddings} as \( E_\theta^{\mathrm{co}} \in \mathbb{R}^{d \cdot m} \), where $m$ is bits per symbol. This embedding yields a vector in $\mathbb{R}^{d\cdot m}$, which we subsequently reshape into $m$ embeddings, each of dimension $d$, effectively providing a distinct embedding of dimension $d$ for each bit within the symbol. The LLR corresponding to each constant embedding should be zero, which is consistent with the behavior of classical puncturing, where punctured bits are treated as unknowns.

To learn this constant embedding, we adopt a variant of the Honda and Yamamoto scheme \cite{honda2013polar}, as used in~\cite{aharoni2023data_arcive} to compute \( P^\theta_{U_i \mid U^{i-1}_1} \). Specifically, In our case the input consists of i.i.d. uniformly distributed bits, \( U_i \sim \text{Bern}(0.5) \),  which implies that \( P_{U_i \mid U^{i-1}_1}(1|u^{i-1}) = 0.5 \) for all $i$. Consequently, we train $E_\theta^{co}(\mathbf{0})$ to satisfy \( H_{\theta}(E_{\theta}^{\mathrm{co}}(\mathbf{0})) = 0 \), which corresponds to an LLR of zero, as required. The detailed training procedure is provided in Section \ref{sec:training}.

\subsection{NPD-Based Communication Systems}
\label{sec:Integrate_NPD}

This section presents NPD-based communication systems for two waveform architectures: OFDM and single-carrier transmission. As depicted in Figure~\ref{fig:combined_systems}, both systems share a common encoder, rate matching, Mapper, and NPD structure, differing only in their waveform processing block. Unlike traditional receivers, which rely on channel estimation and equalization, NPD-based systems replace these components with an NN embedding function. The overall processing chain is consistent with that of a conventional polar decoder, incorporating rate matching followed by SC decoding. However, the NPD uses rate recovery based on constant embeddings and performs SC decoding using NN functions.

Both systems take as input a sequence \( \mathbf{b}^K \) uniformly drawn from \( \{0,1\} \). For CRC-assisted decoding, CRC bits are appended to the input sequence. The bits were encoded using a polar encoder. Specifically, the input vector \( u^N \in \{0,1\}^N \) is constructed such that \( u_i = f_i \) for indices \( i \in \mathcal{F} \) (the frozen set) and \( u_i = b_i \) for indices \( i \in \mathcal{A} \) (the information set). The frozen bits \( f^N \) are defined in~\eqref{eqn:fN} and are generated pseudorandomly using a shared seed, thereby ensuring synchronization between the encoder and decoder. The encoded codeword is computed as \( x^N = u^N G_N \), where \( G_N \) is the polar transform matrix. 

Rate matching is then applied to obtain the codeword \( x^{N_r} \) with \( N_r \leq N \). The codeword is mapped to modulation symbols \( s^{N_r/m} \), where \( m \) is the number of bits per symbol. In the OFDM configuration, this symbol sequence is passed through an OFDM modulator, whereas in the single-carrier case, it is processed through a pulse-shaping filter before transmission over the wireless channel.

At the receiver, the waveform is first demodulated via an OFDM demodulator for OFDM systems or via a matched filter for single-carrier systems. This yields the received symbol sequence \( y^{N_r/m} \).

The decoding process then proceeds through the NPD pipeline, which is identical for both waveform architectures but operates with waveform-specific optimized parameters \( \theta^* \). First, the channel embedding function \( E_{\theta^*} \) maps the received symbols to a latent representation \( e_{0,1}^{N_r} \). Next, Rate Recovery using \( E^{\text{co}}_{\theta^*} \) reconstructs the full-length embedding vector \( e_{0,1}^{N} \),  following the procedure in Section~\ref{sec:rm}. Finally, the SC decoding procedure is applied using the NPD neural network components \( F_{\theta^*}, G_{\theta^*}, H_{\theta^*}\), yielding the estimated bit sequence \( \hat{\mathbf{b}}^K \).

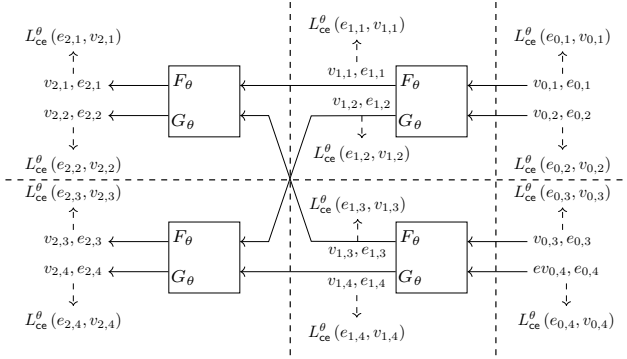
\begin{figure}[h!]
    \centering
    \resizebox{0.5\textwidth}{!}{
    \tikzstyle{gn} = [rectangle,
                    node distance=1.5cm,
                    minimum width=1.5cm, 
                    minimum height=1.25cm,
                    minimum size=1.25cm,
                    text centered, 
                    draw=black,thin,
                    scale=1.0]
\tikzstyle{io} =    [node distance=0.5cm,
                    minimum height=0.25cm,
                    text centered,
                    scale=\scale]

\tikzset{XOR/.style={draw,circle,append after command={
        [shorten >=\pgflinewidth, shorten <=\pgflinewidth,]
        (\tikzlastnode.north) edge (\tikzlastnode.south)
        (\tikzlastnode.east) edge (\tikzlastnode.west)}}}

\tikzset{
    pics/nnops/.style=
    {
        code=
        {
        \node(#1)[gn] at (0,0){} ;
        \node (a) [io, at={($(#1.west)+(0.75em,1em)$)}, scale=1.5, thick] {$\scriptstyle F_\theta$};
        \node (b) [io, at={($(#1.west)+(0.75em,-1em)$)}, scale=1.5, thick] {$\scriptstyle G_\theta$};
        }
    }
} 

\def\scale{0.825}
\begin{tikzpicture}[
node distance = 3cm, 
auto,
]

\pic [] {nnops=Gn11};
\pic [right of=Gn11, xshift=1.cm] {nnops=Gn12};
\pic [below of=Gn11, xshift=0.cm, yshift=0.25cm] {nnops=Gn21};
\pic [right of=Gn21, xshift=1.cm, yshift=0.0cm] {nnops=Gn22};

\draw[->] ([yshift=0.28cm] Gn11.west) -- ++ (-3em, 0) node[anchor=east, pos=1, scale=\scale, pin={[pin edge={black, dashed, ->}, scale=\scale]90:$\celoss{v_{2,1}}{e_{2,1}}$}] {${v_{2,1}},{e_{2,1}}$};
\draw[->] ([yshift=-0.25cm]Gn11.west) -- ++ (-3em, 0) node[anchor=east, pos=1, scale=\scale, pin={[pin edge={black, dashed, ->}, scale=\scale]270:$\celoss{v_{2,2}}{e_{2,2}}$}] {${v_{2,2}},{e_{2,2}}$};
\draw[->] ([yshift=0.28cm] Gn21.west) -- ++ (-3em, 0) node[anchor=east, pos=1, scale=\scale, pin={[pin edge={black, dashed, ->}, scale=\scale]90:$\celoss{v_{2,3}}{e_{2,3}}$}] {${v_{2,3}},{e_{2,3}}$};
\draw[->] ([yshift=-0.25cm]Gn21.west) -- ++ (-3em, 0) node[anchor=east, pos=1, scale=\scale, pin={[pin edge={black, dashed, ->}, scale=\scale]270:$\celoss{v_{2,4}}{e_{2,4}}$}] {${v_{2,4}},{e_{2,4}}$};

\draw[<-] ([yshift=0.28cm]Gn11.east)  -- node[anchor=south, pos=0.75, scale=\scale, pin={[pin edge={black, dashed, ->}, scale=\scale]90:$\celoss{v_{1,1}}{e_{1,1}}$}] {${v_{1,1}},{e_{1,1}}$} ([yshift=0.28cm]Gn12.west);
\draw[<-] ([yshift=0.28cm]Gn21.east)  --++ (0.5,0.0) --++ (0.75,2.215cm) -- node[anchor=south, xshift=-0.0cm, yshift=-0.0cm, scale=\scale, pos=0.6, pin={[pin edge={black, dashed, ->}, scale=\scale]270:$\celoss{v_{1,2}}{e_{1,2}}$}] {${v_{1,2}},{e_{1,2}}$} ([yshift=-0.25cm]Gn12.west);
\draw[<-] ([yshift=-0.25cm]Gn11.east)  --++ (0.5,0.0) --++ (0.75,-2.215cm) -- node[anchor=north, xshift=-0cm, yshift=0.0cm, scale=\scale, pos=0.55, pin={[pin edge={black, dashed, ->}, scale=\scale]90:$\celoss{v_{1,3}}{e_{1,3}}$}] {${v_{1,3}},{e_{1,3}}$} ([yshift=0.28cm]Gn22.west);
\draw[<-] ([yshift=-0.25cm]Gn21.east)  -- node[anchor=north, xshift=0.0cm, yshift=0.00cm, scale=\scale, pos=0.75, pin={[pin edge={black, dashed, ->}, scale=\scale]270:$\celoss{v_{1,4}}{e_{1,4}}$}] {${v_{1,4}},{e_{1,4}}$} ([yshift=-0.25cm]Gn22.west);

\draw[<-] ([yshift=0.28cm] Gn12.east)  -- ++ (3em, 0) node[right, scale=\scale, pin={[pin edge={black, dashed, ->}, scale=\scale]90:$\celoss{v_{0,1}}{e_{0,1}}$}] {${v_{0,1}},{e_{0,1}}$};
\draw[<-] ([yshift=-0.25cm]Gn12.east)  -- ++ (3em, 0) node[right, scale=\scale, pin={[pin edge={black, dashed, ->}, scale=\scale]270:$\celoss{v_{0,2}}{e_{0,2}}$}] {${v_{0,2}},{e_{0,2}}$};
\draw[<-] ([yshift=0.28cm] Gn22.east)  -- ++ (3em, 0) node[right, scale=\scale, pin={[pin edge={black, dashed, ->}, scale=\scale]90:$\celoss{v_{0,3}}{e_{0,3}}$}] {${v_{0,3}},{e_{0,3}}$};
\draw[<-] ([yshift=-0.25cm]Gn22.east)  -- ++ (3em, 0) node[right, scale=\scale, pin={[pin edge={black, dashed, ->}, scale=\scale]270:$\celoss{v_{0,4}}{e_{0,4}}$}] {$e{v_{0,4}},{e_{0,4}}$};

\draw[-,dashed, semithick] ($(Gn11.south)-(3.5, 0.75)$) -- ($(Gn12.south)-(-3.5, 0.75)$);
\draw[-,dashed, semithick] ($(Gn11.east)+(0.88, 1.75)$) -- ($(Gn21.east)+(0.88, -1.75)$);
\draw[-,dashed, semithick] ($(Gn12.east)+(0.5, 1.75)$) -- ($(Gn22.east)+(0.5, -1.75)$);

\end{tikzpicture}
    }
    \caption{A visualization of NSCLoss for $N=4$.}
    \label{fig:nsc-training}
\end{figure}

\subsection{Training Phase}
\label{sec:training}


The NPD parameters are learned through a training phase that consists of an iterative optimization procedure. In each iteration, a batch of channel input-output pairs is used to compute the gradients to update the parameter set $\theta$ via stochastic gradient descent (SGD). The goal of the training phase is to tune $\theta$ such that its outputs would match the true LLRs $\f{L_{U_i|U^{i-1},Y^N}}{y^N,u^{i-1}}, \;i\in[1:N]$ of the synthetic channels, as defined in \eqref{eqn:llr_def}.

\subsubsection{Loss Function}

The training loss, denoted as $\mathsf{NSCLoss}$, was introduced in~\cite{aharoni2023data_arcive}. This loss function is specifically designed to align the NPD outputs with the ground truth LLRs, as defined in~\eqref{eqn:llr_def}. This loss can be computed efficiently using an algorithm with $\log N$ steps, as proposed in~\cite{aharoniCodeRateOptimization2024} and further detailed in Appendix~\ref{sec:NPDloss}. The loss is mathematically defined as:
\begin{equation}\label{eqn:npd_loss}
\mathsf{NSCLoss}(e_{0,1}^N, x^N; \theta) =\frac{1}{(n+1)N} \sum_{l=0}^{n} \sum_{i=1}^{N} L^\theta_\mathsf{ce}(e_{l,i}, v_{l,i}),
\end{equation}
Each term in the loss is a binary cross-entropy, defined as
\begin{equation}\label{eqn:ce_loss}
L^\theta_\mathsf{ce}(e, v) = -v \log H_\theta(e) - (1 - v) \log (1 - H_\theta(e)),
\end{equation}
\noindent where $e_{l,i}$ represents the embedding of the $i$-th bit at the $l$-th decoding depth, and $v_{l,i}$ is the corresponding ground truth bit. Specifically, $e_{0,i}$ denotes the embedding of the channel input $X_i$, and $e_{n,i}$ corresponds to the final embedding of the estimated bit $U_i$. We denote $e_l = (e_{l,1}, \dots, e_{l,N})$ as the vector of embeddings at depth $l$ in the recursive SC decoding structure. Figure~\ref{fig:nsc-training} illustrates the computation of $\mathsf{NSCLoss}$ for $N=4$.

\begin{algorithm}[h!]
    \caption{\gls{npd} for communication Training }
    \label{alg:npd_train}
    \textbf{input:} 
    Dataset $\cD_{B}$, \#of iterations $\mathsf{N_{iters}}$, learning rate $\gamma$ \\
    \textbf{output:} Optimized $\theta^\ast$
    \algrule
    \begin{algorithmic}
    \State Initiate the weights of $E_\theta,E_\theta^{co} F_\theta, G_\theta, H_\theta$
    \For{$\mathsf{N_{iters}}$ iterations} 
        \State Sample $x^N, y^N\sim \cD_{B}$
        \State Compute \( u^N = x^N G_N \)
        \State Compute \( e_0 \leftarrow E_\theta(y^{N/m}, N_0) \)
        \State Duplicate \( E^{co}_\theta(\mathbf{0})\) to $e_0^{co}$
        \State Compute \( \mathcal{L}_X = \text{NSCLoss}(e_0^{co}, x^N; \theta) \) and,
        \State \( \mathcal{L}_Y = \text{NSCLoss}(e_0, x^N; \theta) \) \Comment{ equation \eqref{eqn:npd_loss}}
        \State Update $\theta := \theta - \gamma \nabla_\theta (\mathcal{L}_X + \mathcal{L}_Y)$ 
    \EndFor \\
    \Return $\theta^\ast$ 
    \end{algorithmic}
\end{algorithm}

\subsubsection{Training procedure}
The training procedure is summarized in Algorithm \ref{alg:npd_train} and is detailed below. Let $\mathcal{D}_{M}$ be a dataset consisting of $B$ consecutive blocks, each containing $(x^N,y^{N/m})$, where $x^N$ consists of i.i.d. uniform distributed bits and $y^{N/m}$ is the input of NPD according to the systems described in Section~\ref{sec:Integrate_NPD}, with $N_r = N$. 

At each training iteration, a block $(x^N, y^{N/m}) \sim \mathcal{D}_{B}$ is sampled. The channel outputs $y^{N/m}$ are processed via the embedding function $E_\theta$ to obtain the channel embedding $e_{0,1}^N$. Specifically, for each $i \in [1:N/m]$, the $i$th channel output $y_i$ is mapped to embeddings $e_{0,im - m + 1}^{im} = E_\theta(y_i, N_0)$, and the  all embedding sequence $e_0\in \mathbb{R}^{d \times N}$ is obtained by aggregating all such embeddings across $y^{N/m}$. The constant embedding sequence $e_0^{\mathrm{co}}\in\mathbb{R}^{d\times N}$ is formed by repeating the learned constant channel embedding across all $N$ positions.

These embeddings are used to compute two loss functions: $\mathcal{L}_X = \mathrm{NSCLoss}(e_0, x^N; \theta)$ and $\mathcal{L}_Y = \mathrm{NSCLoss}(e_0^{\mathrm{co}}, x^N; \theta)$. The loss $\mathcal{L}_X$ is designed to optimize the embedding representation of the punctured bit positions as part of the proposed rate-matching strategy, whereas $\mathcal{L}_Y$ aligns the decoder output with the ground-truth LLRs, as defined in Equation~\eqref{eqn:llr_def}. The total loss $\mathcal{L} = \mathcal{L}_X + \mathcal{L}_Y$ is minimized via SGD, which updates the parameters $\theta$. Training proceeds until a predefined number of iterations, $N_{\mathrm{iter}}$ is reached. See \cite{aharoni2023data_arcive} for more details.

\begin{algorithm}[h!]
    \caption{Code Design for \gls{npd} }
    \label{alg:npd_design}
    \textbf{input:} $\theta^*$, $k$ number of information bits \\
    \textbf{output:} $\mathcal{A}, \mathcal{F}$
    \begin{algorithmic}

   \State Generate $x^{N}\sim \text{Bern}(0.5)$
    \State Apply rate matching to obtain $x^{N_r}$ 
    \State  Transmit through the system and receive $y^{N_r/m}$  
    \State Compute $e_{0,1}^{N_r}$ using $E_\theta(y^{N_r/m}, N_0) $ and reshape
    \If{$N\neq N_R$} 
         \State Apply rate recovery using $E_\theta^{co}$ to obtain $e_{0,1}^{N}$ 
    \EndIf
    \State Compute $ [\mathbf{v}_n, \mathbf{e}_n] = \f{\wtilde{\mathsf{NSCLoss}}}{e_{0,1}^N,x^N;\theta^*}$
    
    \State Compute $\f{\what{\sI}_{\theta^\ast}}{U_i;Y^N|U^{i-1}}$ \Comment{equation \ref{eqn:estimate_mi_bern}}
    \State Compute reliability sequence $Q_1^N=\f{\mathsf{SortIndices}}{\what{\sI}_{\theta^\ast}}$
     \State Set $\mathcal{A} = Q_1^{k-1}$, $\mathcal{F} = Q_{k}^N$  \\
     \Return $\mathcal{A}, \mathcal{F}$ 
    \end{algorithmic}
\end{algorithm}

\subsection{Code Design}
The design phase determines the information set $\mathcal{A}$ and frozen set $\mathcal{F}$, which define the polar code structure. This process relies on estimating the mutual information (MI) of the synthesized channels denoted by $\f{\what{\sI}_{\theta^\ast}}{U_i;Y^N|U^{i-1}}$, where $\theta^\ast$ are the optimized parameters of the NPD. A complete definition of the estimated MI is provided in the section.

\subsubsection{Estimate MI}
To compute the MI, we followed the methodology presented in~\cite{aharoniCodeRateOptimization2024}. Specifically, we utilized Algorithm~$\widetilde{\mathsf{NSCLoss}}$, which modifies the standard $\mathsf{NSCLoss}$. Rather than computing the loss $\mathsf{NSCLoss}(e_{0,1}^N, x^N; \theta)$, this algorithm outputs both the bits $v_n = u^N$ and their corresponding embeddings $e_n$. The algorithm is formally defined as follows:
\begin{equation}
    [\mathbf{v}_n, \mathbf{e}_n] = \f{\wtilde{\mathsf{NSCLoss}}}{e_{0,1}^N,x^N;\theta^\prime},
\end{equation}
Then, the estimated \gls{mi} of the synthetic channels is computed for all $i\in[1:N]$ assuming $P_x\sim \text{Bern}(0.5)$
\begin{align}
    \f{\what{\sI}_{\theta^\ast}}{U_i;Y^N|U^{i-1}} &= \notag \\ 
    &\hspace{-2.65cm}\frac{1}{B}\hspace{0cm}\sum_{x^N,y^{N_r/m}\in\mathcal{D}_{B}}\hspace{-0.65cm} 
    &\hspace{-2.4cm}1-\hspace{-0.1cm} L_{ce}^\theta(e_{n,i},v_{n,i}), 
\label{eqn:estimate_mi_bern}
\end{align}

\subsubsection{Design Procedure}

The NPD-based polar code design procedure is outlined in Algorithm~\ref{alg:npd_design}. It begins by generating a random input \( x^N \sim \text{Bern}(0.5) \). If required, rate matching is applied to produce a shortened codeword \( x^{N_r} \), which is modulated and transmitted. The received signal \( y^{N_r/m} \) is processed by the embedding function \( E_\theta \), yielding the sequence \( e_{0,1}^{N_r} \). If \( N \ne N_r \), the full embedding \( e_{0,1}^N \) is reconstructed using the constant embedding function \( E_\theta^{\mathrm{co}} \).

Then apply \( \widetilde{\mathsf{NSCLoss}} \) to obtain \( \mathbf{v}_n \) and \( \mathbf{e}_n \). These are used to estimate the MI for each bit channel using Equation~\eqref{eqn:estimate_mi_bern}. The MI values are sorted to form a reliability sequence \( Q_1^N \), from which the top \( k = \lfloor RN \rfloor \) indices define the information set \( \cA \); the remaining indices form the frozen set \( \cF = [1:N] \setminus \cA \).

\begin{table}[h!]
\centering
\caption{Parameters Used for Training and Evaluation}
\label{tab:parameters_results}
\begin{tabular}{|l|l|}
\hline
\textbf{Parameter} & \textbf{Value} \\
\hline
\multicolumn{2}{|c|}{\textbf{Training and Model Configuration}} \\
\hline
Power delay profile            & TDL-C (300\,ns) \\
Noise variance range  (uniform)     & --5 to 15\,dB \\
Velocity range (uniform)                & 0 to 35\,m/s \\
Batch size                     & 100 \\
Training iterations            & $10^6$ \\
code length                   & 1024 \\
Embedding dimension ($d$)      & 128 \\
Neurons units ($h$)      & 300 \\
Learning rate                  & $10^{-3}$ \\
Optimization algorithm         & Adam \\
\hline
\multicolumn{2}{|c|}{\textbf{OFDM Configuration}} \\
\hline
Carrier frequency              & 3.5\,GHz \\
Subcarrier spacing             & 15\,kHz \\
\hline
\multicolumn{2}{|c|}{\textbf{Classic System Configuration}} \\
\hline
OFDM channel estimation     & Least Squares \\
OFDM equalizer          & Linear MMSE \\
single-carrier channel estimation          & Perfect \\
single-carrier equalizer                   & Zero Forcing  \\
5G Polar configuration & Uplink \\ 
Cyclic Prefix & $4.69\mu s$ \\
\hline
\multicolumn{2}{|c|}{\textbf{Evaluation Configuration}} \\
\hline
Power delay profile            & TDL-A \\
List decoder size              & 16 \\
CRC length                     & 11 \\
\hline
\end{tabular}
\end{table}

\section{EXPERIMENTS AND INSIGHTS}\label{sec:exp}
This section presents a comprehensive evaluation of the NPD for communication by benchmarking its performance against standardized 5G polar code decoders. The assessment was conducted under realistic and challenging channel conditions. We examine a wide range of parameters that influence modern communication systems, including delay spreads, waveform architectures (single-carrier and OFDM), modulation schemes (BPSK and QPSK), code lengths, rate matching techniques, Doppler effects, and nonlinear distortions introduced by power amplifiers.

\begin{figure}[h!]
\centering

\begin{subfigure}[b]{0.2\textwidth}
    \centering
    \begin{tikzpicture}[scale=0.22]
        \foreach \x in {0,...,9} {
            \foreach \y in {0,...,9} {
                \filldraw[white] (\x,\y) rectangle ++(1,1);
                \draw (\x,\y) rectangle ++(1,1);
            }
        }
        \foreach \y in {0,...,9} {
            \filldraw[black] (2,\y) rectangle ++(1,1);
        }
        \node at (5,-1.5) {OFDM symbol};
        \node[rotate=90] at (-1.5,5) {Subcarrier};
    \end{tikzpicture}
    \caption{Pilot pattern "$1P$"}
\end{subfigure}
\hspace{1cm}
\begin{subfigure}[b]{0.2\textwidth}
    \centering
        \begin{tikzpicture}[scale=0.22]
        \foreach \x in {0,...,9} {
            \foreach \y in {0,...,9} {
                \filldraw[white] (\x,\y) rectangle ++(1,1);
                \draw (\x,\y) rectangle ++(1,1);
            }
        }
        \foreach \y in {0,...,9} {
            \filldraw[black] (2,\y) rectangle ++(1,1);
            \filldraw[black] (7,\y) rectangle ++(1,1);
        }
        \node at (5,-1.5) {OFDM symbol};
        \node[rotate=90] at (-1.5,5) {Subcarrier};
    \end{tikzpicture}
    \caption{Pilot pattern "$2P$"}
\end{subfigure}

\vspace{0.8cm} 

\begin{subfigure}[b]{0.2\textwidth}
    \centering
    \begin{tikzpicture}[scale=0.22]
        \foreach \x in {0,...,9} {
            \foreach \y in {0,...,9} {
                \filldraw[white] (\x,\y) rectangle ++(1,1);
                \draw (\x,\y) rectangle ++(1,1);
            }
        }
        \foreach \y in {1,3,5,7,9} {
            \filldraw[black] (2,\y) rectangle ++(1,1);
        }
        \node at (5,-1.5) {OFDM symbol};
        \node[rotate=90] at (-1.5,5) {Subcarrier};
    \end{tikzpicture}
    \caption{Pilot pattern "$1P_2$"}
\end{subfigure}
\hspace{1cm}
\begin{subfigure}[b]{0.2\textwidth}
    \centering
    \begin{tikzpicture}[scale=0.22]
        \foreach \x in {0,...,9} {
            \foreach \y in {0,...,9} {
                \filldraw[white] (\x,\y) rectangle ++(1,1);
                \draw (\x,\y) rectangle ++(1,1);
            }
        }
        \foreach \y in {0,1,...,9} {
            \ifodd\y
                \filldraw[black] (2,\y) rectangle ++(1,1);
                \filldraw[black] (7,\y) rectangle ++(1,1);
            \fi
        }
        \node at (5,-1.5) {OFDM symbol};
        \node[rotate=90] at (-1.5,5) {Subcarrier};
    \end{tikzpicture}
    \caption{Pilot pattern "$2P_2$"}
\end{subfigure}

\caption{Pilot patterns used in simulations.}
\label{fig:pilot_patterns}
\end{figure}
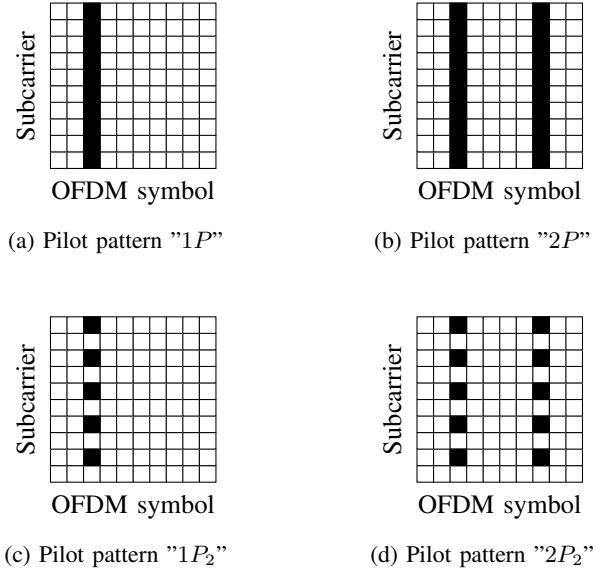

\subsection{Setup}

The experiments were conducted on single-input single-output (SISO) systems using both OFDM and single-carrier transmissions, as outlined in Section~\ref{sec:Integrate_NPD}. The communication channel was modeled according to the multipath fading channel described in Section~\ref{sec:communication:model}, using the 3GPP TDL profiles~\cite{3gpp38901}. Our implementation was developed in Python using TensorFlow \cite{abadi2016tensorflow} for neural network modeling and the Sionna library for communication system simulations \cite{hoydis2022sionna}. Table \ref{tab:parameters_results} lists the parameters used for training and evaluation.

 \subsubsection*{\textbf{Model Evaluation}} Throughout the experiments, we compared the proposed NPD against the standard 5G uplink polar code \cite{3gpp38212} that relies on conventional equalization and channel estimation, both employing CRC-assisted list decoding (CA-SCL). Training was conducted using TDL-C power delay profiles, and evaluation was performed on TDL-A profiles to avoid overfitting to a specific channel model. 
 
 A single model per modulation scheme and system waveform (OFDM or SC) was trained and used across all channel conditions and code sizes in the evaluation, demonstrating the generalization capability of the model without requiring retraining or fine-tuning.

\subsubsection*{\textbf{Frame Structure}} In OFDM experiments with BPSK and a code size of 1024, the system uses eight OFDM symbols and 128 subcarriers. For smaller code sizes (864, 512, 432, 256, and 216), the number of subcarriers was set to 108, 64, 54, 32, and 27, respectively. In the QPSK experiments, 64 subcarriers were used for a code size of 1024 and 32 for 512. In the 5G polar code OFDM-based system, the CP is configured to maintain an approximate 7\% overhead. For instance, with a code size of 1024 and 128 subcarriers, a CP length of 9 yields \( T_{\text{cp}} = 4.68 \mu \,\text{s} \), closely matching the 3GPP normal CP of \( 4.69 \mu \,\text{s} \) at 15kHz spacing.

The 5G polar decoder was evaluated under four pilot configurations, as shown in Figure~\ref{fig:pilot_patterns}. In the "$1P$" and "$2P$" settings, the pilots occupy all subcarriers in one or two OFDM symbols, respectively. In the "$1P_2$" and "$2P_2$" settings, pilots are placed on every second subcarrier within the designated symbols. Unless stated otherwise, the $2P$ configuration was used by default. All pilots were randomly generated using QPSK modulation. Note that when comparing 5G polar and NPD at the same code length, extra OFDM symbols were added for pilots in the 5G decoder based on the chosen configuration.

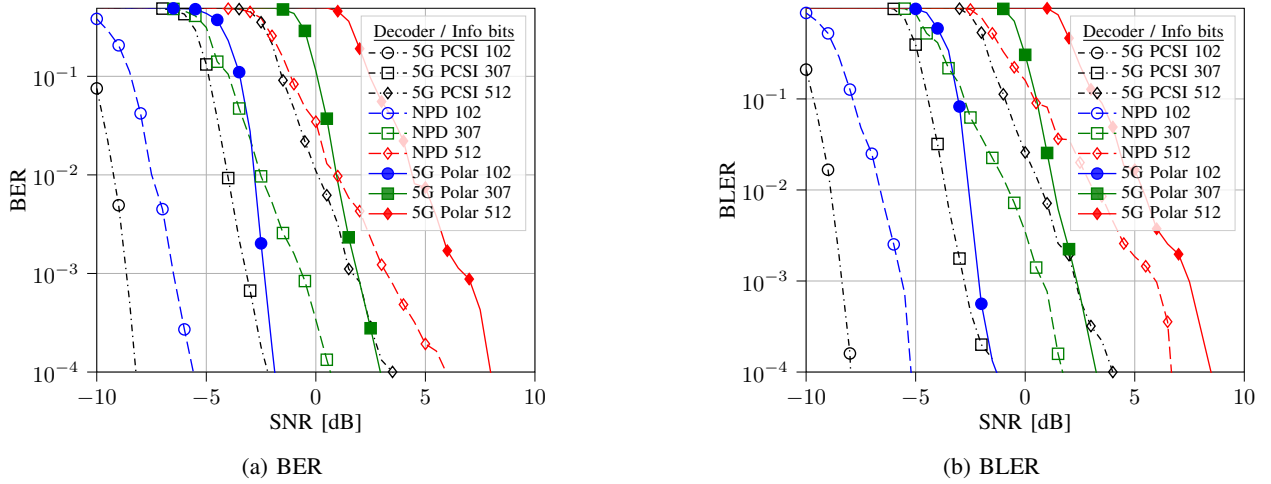
\begin{figure*}[t!]
\centering

\begin{subfigure}[b]{0.48\textwidth}
    \centering
    \scalebox{0.85}{
\begin{tikzpicture}

\definecolor{darkgray176}{RGB}{176,176,176}
\definecolor{green01270}{RGB}{0,127,0}
\definecolor{lightgray204}{RGB}{204,204,204}

\begin{axis}[
legend cell align={left},
legend style={fill opacity=0.8, draw opacity=1,   font=\footnotesize,
text opacity=1, draw=lightgray204,  row sep=-2pt },%
log basis y={10},
tick align=outside,
tick pos=left,
x grid style={darkgray176},
xlabel={SNR [dB]},
xmajorgrids,
xmin=-10, xmax=10,
xtick style={color=black},
y grid style={darkgray176},
ylabel={BER},
ymajorgrids,
ymin=0.0001, ymax=0.49,
ymode=log,
ytick style={color=black},
ytick={0.0001,0.001,0.01,0.1,1,10},
yticklabels={
  \(\displaystyle {10^{-4}}\),
  \(\displaystyle {10^{-3}}\),
  \(\displaystyle {10^{-2}}\),
  \(\displaystyle {10^{-1}}\),
  \(\displaystyle {10^{0}}\),
  \(\displaystyle {10^{1}}\)
}
]
\addlegendimage{empty legend}
\addlegendentry{\hspace{-.6cm}\underline{Decoder / Info bits}}

\addplot [semithick, black, dash dot, mark=o, mark size=2.5, mark repeat=2, mark options={solid,fill opacity=0}]
table {%
-10 0.07544117647058823
-9.5 0.02235294117647059
-9 0.004923747276688453
-8.5 0.0005140774258421317
-8 0.00002980392156862745
-7.5 0
};
\addlegendentry{5G PCSI 102}

\addplot [semithick, black, dash dot, mark=square, mark size=2.5, mark repeat=2, mark options={solid,fill opacity=0}]
table {%
-10	0.502557003257329
-9.5	0.49741042345276900
-9	0.4989413680781760
-8.5	0.4967100977198700
-8	0.4942019543973940
-7.5	0.4916775244299670
-7	0.48657980456026100
-6.5	0.4667915309446250
-6	0.42750814332247600
-5.5	0.29563517915309400
-5	0.1319543973941370
-4.5	0.03736156351791530
-4	0.009286644951140060
-3.5	0.002204361988386910
-3	0.0006692853036980260
-2.5	0.0001793485342019540
-2	0.00006925081433224760
-1.5	0.00005322475570032570
-1	0
-0.5	0
0	0
};
\addlegendentry{5G PCSI 307}

\addplot [semithick, black, dash dot, mark=diamond, mark size=2.5, mark repeat=2, mark options={solid,fill opacity=0}]
table {%
-10	0.49984
-9.5	0.5013
-9	0.49855
-8.5	0.49856
-8	0.50009
-7.5	0.49625
-7	0.49535
-6.5	0.49708
-6	0.49696
-5.5	0.49753
-5	0.49518
-4.5	0.49444
-4	0.49474
-3.5	0.48034
-3	0.45405
-2.5	0.35372
-2	0.21416
-1.5	0.091074
-1	0.046948
-0.5	0.021823
0	0.010999
0.5	0.0062207
1	0.0031836
1.5	0.0011099
2	0.00082181
2.5	0.00028647
3	0.00013191
3.5	0.00010023
4	0.000043984
};
\addlegendentry{5G PCSI 512}

\addplot [semithick, blue, dash pattern=on 5.55pt off 2.4pt, mark=o, mark size=2.5, mark repeat=2, mark options={solid,fill opacity=0}]
table {%
-10 0.382433628318584
-9.5 0.292345132743363
-9 0.206238938053097
-8.5 0.111283185840708
-8 0.0420648967551622
-7.5 0.010117994100295
-7 0.00449316170555109
-6.5 0.000996245642263341
-6 0.000270850093858943
-5.5 7.97234190768596e-05
-5 0
-4.5 0
-4 0
-3.5 0
-3 0
-2.5 0
-2 0
-1.5 0
-1 0
-0.5 0
0 0
0.5 0
1 0
1.5 0
2 0
2.5 0
3 0
3.5 0
4 0
4.5 0
5 0
5.5 0
6 0
6.5 0
7 0
7.5 0
8 0
8.5 0
9 0
9.5 0
10 0
10.5 0
11 0
11.5 0
12 0
12.5 0
13 0
13.5 0
14 0
14.5 0
15 0
15.5 0
16 0
16.5 0
17 0
17.5 0
18 0
18.5 0
19 0
19.5 0
};
\addlegendentry{NPD 102}
\addplot [semithick, green01270, dash pattern=on 5.55pt off 2.4pt, mark=square, mark size=2.5, mark repeat=2, mark options={solid,fill opacity=0}]
table {%
-10 0.496305031446541
-9.5 0.50124213836478
-9 0.500660377358491
-8.5 0.497877358490566
-8 0.500487421383648
-7.5 0.492798742138365
-7 0.495440251572327
-6.5 0.482547169811321
-6 0.459496855345912
-5.5 0.410566037735849
-5 0.313176100628931
-4.5 0.140754716981132
-4 0.105
-3.5 0.0471540880503145
-3 0.0231682389937107
-2.5 0.00966588050314465
-2 0.00525157232704403
-1.5 0.00257206498951782
-1 0.00159053293611387
-0.5 0.000837700908455625
0 0.000338923829489867
0.5 0.000133294466822133
1 5.53690627773276e-05
1.5 0
2 0
2.5 0
3 0
3.5 0
4 0
4.5 0
5 0
5.5 0
6 0
6.5 0
7 0
7.5 0
8 0
8.5 0
9 0
9.5 0
10 0
10.5 0
11 0
11.5 0
12 0
12.5 0
13 0
13.5 0
14 0
14.5 0
15 0
15.5 0
16 0
16.5 0
17 0
17.5 0
18 0
18.5 0
19 0
19.5 0
};
\addlegendentry{NPD 307}
\addplot [semithick, red, dash pattern=on 5.55pt off 2.4pt, mark=diamond, mark size=2.5, mark repeat=2, mark options={solid,fill opacity=0}]
table {%
-10 0.49802103250478
-9.5 0.500650095602294
-9 0.500325047801147
-8.5 0.499923518164436
-8 0.498126195028681
-7.5 0.499789674952199
-7 0.499875717017208
-6.5 0.497437858508604
-6 0.495697896749522
-5.5 0.494579349904398
-5 0.494579349904398
-4.5 0.491739961759082
-4 0.485143403441683
-3.5 0.478881453154876
-3 0.449847036328872
-2.5 0.390401529636711
-2 0.257160611854685
-1.5 0.154980879541109
-1 0.0832122370936902
-0.5 0.0484082217973231
0 0.0346414913957935
0.5 0.0130146590184831
1 0.00971797323135755
1.5 0.00585768915596831
2 0.0042925430210325
2.5 0.00227459920576555
3 0.00122530274059911
3.5 0.000789356277884003
4 0.000483921432296193
4.5 0.000320612643157044
5 0.000192833368741591
5.5 0.00016324647605496
6 9.37865919634784e-05
6.5 0
7 0
7.5 0
8 0
8.5 0
9 0
9.5 0
10 0
10.5 0
11 0
11.5 0
12 0
12.5 0
13 0
13.5 0
14 0
14.5 0
15 0
15.5 0
16 0
16.5 0
17 0
17.5 0
18 0
18.5 0
19 0
19.5 0
};
\addlegendentry{NPD 512}
\addplot [semithick, blue, mark=*, mark size=2.5, mark repeat=2, mark options={solid}]
table {%
-10 0.498578431372549
-9.5 0.500686274509804
-9 0.499656862745098
-8.5 0.494460784313725
-8 0.494117647058824
-7.5 0.490833333333333
-7 0.495049019607843
-6.5 0.488921568627451
-6 0.486617647058824
-5.5 0.479607843137255
-5 0.434705882352941
-4.5 0.373578431372549
-4 0.230196078431373
-3.5 0.110196078431373
-3 0.0276225490196078
-2.5 0.00201907790143084
-2 0.00017724168319013
-1.5 0.000017724168319013
-1 0 0.000001775490196078
-0.5 0
0 0
0.5 0
1 0
1.5 0
2 0
2.5 0
3 0
3.5 0
4 0
4.5 0
5 0
5.5 0
6 0
6.5 0
7 0
7.5 0
8 0
8.5 0
9 0
9.5 0
10 0
10.5 0
11 0
11.5 0
12 0
12.5 0
13 0
13.5 0
14 0
14.5 0
15 0
15.5 0
16 0
16.5 0
17 0
17.5 0
18 0
18.5 0
19 0
19.5 0
};
\addlegendentry{5G Polar 102}
\addplot [semithick, green01270, mark=square*, mark size=2.5, mark repeat=2, mark options={solid}]
table {%
-10 0.49885993485342
-9.5 0.502817589576547
-9 0.500472312703583
-8.5 0.500325732899023
-8 0.497524429967427
-7.5 0.499657980456026
-7 0.497019543973941
-6.5 0.501742671009772
-6 0.500635179153094
-5.5 0.50071661237785
-5 0.503517915309446
-4.5 0.500309446254072
-4 0.496579804560261
-3.5 0.497231270358306
-3 0.497182410423453
-2.5 0.492231270358306
-2 0.493827361563518
-1.5 0.476644951140065
-1 0.434267100977199
-0.5 0.28900651465798
0 0.113811074918567
0.5 0.0371769815418024
1 0.00846742671009772
1.5 0.00232520263616393
2 0.000811046345505457
2.5 0.000278065788870028
3 0.0000878065788870028
3.5 0
4 0
4.5 0
5 0
5.5 0
6 0
6.5 0
7 0
7.5 0
8 0
8.5 0
9 0
9.5 0
10 0
10.5 0
11 0
11.5 0
12 0
12.5 0
13 0
13.5 0
14 0
14.5 0
15 0
15.5 0
16 0
16.5 0
17 0
17.5 0
18 0
18.5 0
19 0
19.5 0
};
\addlegendentry{5G Polar 307}
\addplot [semithick, red, mark=diamond*, mark size=2.5, mark repeat=2, mark options={solid}]
table {%
-10 0.501162109375
-9.5 0.498701171875
-9 0.499921875
-8.5 0.5003125
-8 0.498056640625
-7.5 0.499228515625
-7 0.499775390625
-6.5 0.503251953125
-6 0.501728515625
-5.5 0.495947265625
-5 0.499716796875
-4.5 0.49724609375
-4 0.502197265625
-3.5 0.499482421875
-3 0.498134765625
-2.5 0.497607421875
-2 0.49958984375
-1.5 0.499033203125
-1 0.500478515625
-0.5 0.495400390625
0 0.49876953125
0.5 0.4940625
1 0.45791015625
1.5 0.3620703125
2 0.190634765625
2.5 0.09576171875
3 0.05548828125
3.5 0.037900390625
4 0.0220947265625
4.5 0.00814592633928571
5 0.00744596354166667
5.5 0.0037890625
6 0.00170209099264706
6.5 0.00113420758928571
7 0.000877522145669291
7.5 0.000428695678710937
8 0.0000928695678710937
8.5 0
9 0
9.5 0
10 0
10.5 0
11 0
11.5 0
12 0
12.5 0
13 0
13.5 0
14 0
14.5 0
15 0
15.5 0
16 0
16.5 0
17 0
17.5 0
18 0
18.5 0
19 0
19.5 0
};
\addlegendentry{5G Polar 512}
\end{axis}

\end{tikzpicture}}
    \caption{BER}
    \label{fig:ber_combined}
\end{subfigure}
\hfill
\begin{subfigure}[b]{0.48\textwidth}
    \centering
    \scalebox{0.85}{
\begin{tikzpicture}

\definecolor{darkgray176}{RGB}{176,176,176}
\definecolor{green01270}{RGB}{0,127,0}
\definecolor{lightgray204}{RGB}{204,204,204}

\begin{axis}[
legend cell align={left},
legend style={fill opacity=0.8, draw opacity=1, text opacity=1,  font=\footnotesize, draw=lightgray204 ,  row sep=-2pt},
log basis y={10},
tick align=outside,
tick pos=left,
x grid style={darkgray176},
xlabel={SNR [dB]},
xmajorgrids,
xmin=-10, xmax=10,
xtick style={color=black},
y grid style={darkgray176},
ylabel={BLER},
ymajorgrids,
ymin=0.0001, ymax=0.99,
ymode=log,
ytick style={color=black},
ytick={0.0001,0.001,0.01,0.1,1,10},
yticklabels={
  \(\displaystyle {10^{-4}}\),
  \(\displaystyle {10^{-3}}\),
  \(\displaystyle {10^{-2}}\),
  \(\displaystyle {10^{-1}}\),
  \(\displaystyle {10^{0}}\),
  \(\displaystyle {10^{1}}\)
}
]
\addlegendimage{empty legend}
\addlegendentry{\hspace{-.6cm}\underline{Decoder / Info bits}}

\addplot [semithick, black, dash dot, mark=o, mark size=2.5, mark repeat=2, mark options={solid,fill opacity=0}]
table {%
-10 0.21
-9.5 0.06833333333333330
-9 0.016666666666666700
-8.5 0.0019230769230769200
-8 0.00016
-7.5 0.0000001
};
\addlegendentry{5G PCSI 102}

\addplot [semithick, black, dash dot, mark=square, mark size=2.5, mark repeat=2, mark options={solid,fill opacity=0}]
table {%
-10	1
-9.5	1
-9	1
-8.5	1
-8	1
-7.5	1
-7	1
-6.5	1
-6	0.975
-5.5	0.76
-5	0.395
-4.5	0.1175
-4	0.032
-3.5	0.006521739130434780
-3	0.0017647058823529400
-2.5	0.00046
-2	0.0002
-1.5	0.00014
-1	0
-0.5	0
};

\addlegendentry{5G PCSI 307}

\addplot [semithick, black, dash dot, mark=diamond, mark size=2.5, mark repeat=2, mark options={solid,fill opacity=0}]
table {%
-10	1
-9.5	1
-9	1
-8.5	1
-8	1
-7.5	1
-7	1
-6.5	1
-6	1
-5.5	1
-5	1
-4.5	1
-4	1
-3.5	1
-3	0.98
-2.5	0.825
-2	0.535
-1.5	0.22
-1	0.1125
-0.5	0.051667
0	0.025833
0.5	0.014091
1	0.0071429
1.5	0.0025833
2	0.0019231
2.5	0.00062762
3	0.00032
3.5	0.00022
4	0.0001
};
\addlegendentry{5G PCSI 512}

\addplot [semithick, blue, dash pattern=on 5.55pt off 2.4pt, mark=o, mark size=2.5, mark repeat=2, mark options={solid,fill opacity=0}]
table {%
-10 0.88
-9.5 0.715
-9 0.525
-8.5 0.29
-8 0.126666666666667
-7.5 0.0483333333333333
-7 0.025
-6.5 0.00757575757575758
-6 0.00252525252525253
-5.5 0.000728862973760933
-5 0.0000239
-4.5 0.0000465
-4 0
-3.5 0
-3 0
-2.5 0
-2 0
-1.5 0
-1 0
-0.5 0
0 0
0.5 0
1 0
1.5 0
2 0
2.5 0
3 0
3.5 0
4 0
4.5 0
5 0
5.5 0
6 0
6.5 0
7 0
7.5 0
8 0
8.5 0
9 0
9.5 0
10 0
10.5 0
11 0
11.5 0
12 0
12.5 0
13 0
13.5 0
14 0
14.5 0
15 0
15.5 0
16 0
16.5 0
17 0
17.5 0
18 0
18.5 0
19 0
19.5 0
};
\addlegendentry{NPD 102}
\addplot [semithick, green01270, dash pattern=on 5.55pt off 2.4pt, mark=square, mark size=2.5, mark repeat=2, mark options={solid,fill opacity=0}]
table {%
-10 1
-9.5 1
-9 1
-8.5 1
-8 1
-7.5 1
-7 1
-6.5 1
-6 0.995
-5.5 0.98
-5 0.865
-4.5 0.525
-4 0.415
-3.5 0.2175
-3 0.145
-2.5 0.0625
-2 0.0378571428571429
-1.5 0.0225
-1 0.0134210526315789
-0.5 0.00722222222222222
0 0.00347222222222222
0.5 0.00140449438202247
1 0.000766871165644172
1.5 0.0001586
2 0.000052325
2.5 0
3 0
3.5 0
4 0
4.5 0
5 0
5.5 0
6 0
6.5 0
7 0
7.5 0
8 0
8.5 0
9 0
9.5 0
10 0
10.5 0
11 0
11.5 0
12 0
12.5 0
13 0
13.5 0
14 0
14.5 0
15 0
15.5 0
16 0
16.5 0
17 0
17.5 0
18 0
18.5 0
19 0
19.5 0
};
\addlegendentry{NPD 307}
\addplot [semithick, red, dash pattern=on 5.55pt off 2.4pt, mark=diamond, mark size=2.5, mark repeat=2, mark options={solid,fill opacity=0}]
table {%
-10 1
-9.5 1
-9 1
-8.5 1
-8 1
-7.5 1
-7 1
-6.5 1
-6 1
-5.5 1
-5 1
-4.5 1
-4 1
-3.5 1
-3 1
-2.5 0.96
-2 0.785
-1.5 0.525
-1 0.33
-0.5 0.2225
0 0.16
0.5 0.09
1 0.08125
1.5 0.0364285714285714
2 0.0357142857142857
2.5 0.02
3 0.01125
3.5 0.0085
4 0.00454545454545455
4.5 0.00257731958762887
5 0.00185185185185185
5.5 0.00145348837209302
6 0.000968992248062015
6.5 0.0003567158628
7 0.00000949872
7.5 0
8 0
8.5 0
9 0
9.5 0
10 0
10.5 0
11 0
11.5 0
12 0
12.5 0
13 0
13.5 0
14 0
14.5 0
15 0
15.5 0
16 0
16.5 0
17 0
17.5 0
18 0
18.5 0
19 0
19.5 0
};
\addlegendentry{NPD 512}
\addplot [semithick, blue, mark=*, mark size=2.5, mark repeat=2, mark options={solid}]
table {%
-10 1
-9.5 1
-9 1
-8.5 1
-8 1
-7.5 1
-7 1
-6.5 1
-6 1
-5.5 0.995
-5 0.975
-4.5 0.885
-4 0.595
-3.5 0.34
-3 0.0825
-2.5 0.00675675675675676
-2 0.000561797752808989
-1.5 0.000132468
-1 0.0000632468
-0.5 0
0 0
0.5 0
1 0
1.5 0
2 0
2.5 0
3 0
3.5 0
4 0
4.5 0
5 0
5.5 0
6 0
6.5 0
7 0
7.5 0
8 0
8.5 0
9 0
9.5 0
10 0
10.5 0
11 0
11.5 0
12 0
12.5 0
13 0
13.5 0
14 0
14.5 0
15 0
15.5 0
16 0
16.5 0
17 0
17.5 0
18 0
18.5 0
19 0
19.5 0
};
\addlegendentry{5G Polar 102}
\addplot [semithick, green01270, mark=square*, mark size=2.5, mark repeat=2, mark options={solid}]
table {%
-10 1
-9.5 1
-9 1
-8.5 1
-8 1
-7.5 1
-7 1
-6.5 1
-6 1
-5.5 1
-5 1
-4.5 1
-4 1
-3.5 1
-3 1
-2.5 1
-2 1
-1.5 1
-1 0.975
-0.5 0.715
0 0.305
0.5 0.103333333333333
1 0.0255
1.5 0.00616279069767442
2 0.00223684210526316
2.5 0.000704225352112676
3 0 0.00010422533232643
3.5 0.000050234
4 0
4.5 0
5 0
5.5 0
6 0
6.5 0
7 0
7.5 0
8 0
8.5 0
9 0
9.5 0
10 0
10.5 0
11 0
11.5 0
12 0
12.5 0
13 0
13.5 0
14 0
14.5 0
15 0
15.5 0
16 0
16.5 0
17 0
17.5 0
18 0
18.5 0
19 0
19.5 0
};
\addlegendentry{5G Polar 307}
\addplot [semithick, red, mark=diamond*, mark size=2.5, mark repeat=2, mark options={solid}]
table {%
-10 1
-9.5 1
-9 1
-8.5 1
-8 1
-7.5 1
-7 1
-6.5 1
-6 1
-5.5 1
-5 1
-4.5 1
-4 1
-3.5 1
-3 1
-2.5 1
-2 1
-1.5 1
-1 1
-0.5 1
0 1
0.5 1
1 0.99
1.5 0.84
2 0.465
2.5 0.2325
3 0.1275
3.5 0.09
4 0.0491666666666667
4.5 0.0185714285714286
5 0.0166666666666667
5.5 0.00833333333333333
6 0.00375
6.5 0.00255102040816327
7 0.00196850393700787
7.5 0.0009765625
8 0 0.00027562
8.5 0.0000955762
9 0
9.5 0
10 0
10.5 0
11 0
11.5 0
12 0
12.5 0
13 0
13.5 0
14 0
14.5 0
15 0
15.5 0
16 0
16.5 0
17 0
17.5 0
18 0
18.5 0
19 0
19.5 0
};
\addlegendentry{5G Polar 512}
\end{axis}

\end{tikzpicture}}
    \caption{BLER}
    \label{fig:bler_combined}
\end{subfigure}

\caption{BER and BLER over a TDL-A channel with 100\,ns delay spread and \(0\text{--}8\,\text{m/s}\) mobility. The system uses OFDM with BPSK modulation and code length \(N = 1024\). The NPD is compared to the 5G Polar code with LS estimation and Perfect CSI. Results are shown for varying information bit lengths.}
\label{fig:ber_bler_combined}
\end{figure*}

\subsubsection*{\textbf{Design}} The information set $\mathcal{A}$ for is found using the design phase of NPD for each SNR point, modulation, and code length. It assumes knowledge of the design condition at the encoder for selecting the design. In contrast, the 5G decoder relies on a static reliability sequence as specified in the 3GPP standard \cite{3gpp38212}. This provides a deployment advantage for the standard 5G decoder.

\subsubsection*{\textbf{Training Configuration}}

This training configuration was carefully designed to balance performance and robustness while maintaining a manageable computational complexity. A thorough exploration of the training hyperparameters was conducted to determine an effective setup for the proposed NPD. This investigation included variations in the embedding dimension, number of neurons, batch size and learning rate. 

To develop a robust model, training was performed under challenging conditions, including a delay spread of 300ns, which resulted in a model capable of generalizing across a wide range of delay spreads. To ensure resilience to varying SNRs, the training dataset was generated using noise variances sampled uniformly from the range \([-5, 10]\) dB. Furthermore, to improve robustness to Doppler effects, the training channels included user velocities uniformly distributed between 0 and 35m/s.

\subsection{Performance over Channel with Nominal Delay Spread}

This experiment compares the performance of the proposed NPD and the 5G polar decoder over a TDL-A channel with a nominal delay spread of $100ns$. To this end, we evaluate both decoders within an OFDM system employing BPSK modulation, with user mobility uniformly distributed in the range \([0, 8] \text{m/s}\). The results include comparisons using LS estimation and Perfect Channel State Information (PCSI) for the 5G polar code.

\textbf{Insight}: The experiment shows that the proposed NPD consistently outperforms the practical 5G polar decoder across all evaluated code rates over a TDL-A channel with nominal delay. The performance advantage becomes increasingly pronounced as the code rate decreases. Notably, the NPD achieves this gain without using pilot symbols and a CP, thereby reducing the transmission overhead. One possible reason for this gain is that the 5G Polar uses least-squares (LS) channel estimation from the pilots, whereas the NPD can learn to extract better channel estimates. This possibility is also supported by our comparison with perfect channel estimation (PCSI) in Figure \ref{fig:ber_combined}.

This insight is obtained from the Figures. \ref{fig:ber_combined} and~\ref{fig:bler_combined}, which illustrate the BER and BLER, respectively, as functions of SNR for various numbers of transmitted information bits. From these figures, we observe that the NPD consistently achieves a lower BLER and BER than the 5G decoder for all SNR values and information bits.

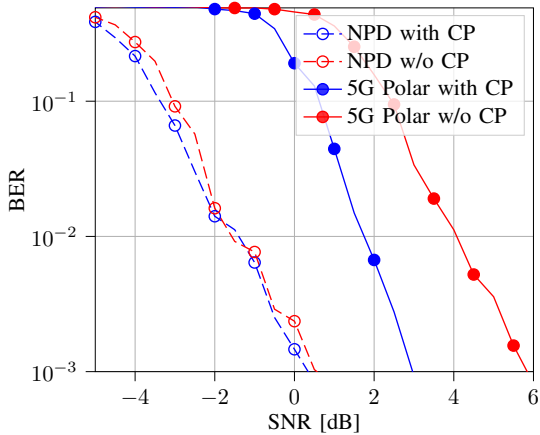
\begin{figure}[h!]
\centering
\scalebox{.85}{
\begin{tikzpicture}

\definecolor{darkgray176}{RGB}{176,176,176}
\definecolor{lightgray204}{RGB}{204,204,204}

\begin{axis}[
legend cell align={left},
legend style={fill opacity=0.8, draw opacity=1, text opacity=1, draw=lightgray204},
log basis y={10},
tick align=outside,
tick pos=left,
x grid style={darkgray176},
xlabel={SNR [dB]},
xmajorgrids,
xmin=-5, xmax=6,
xtick style={color=black},
y grid style={darkgray176},
ylabel={BER},
ymajorgrids,
ymin=0.001, ymax=0.49,
ymode=log,
ytick style={color=black},
ytick={0.0001,0.001,0.01,0.1,1,10},
yticklabels={
  \(\displaystyle {10^{-4}}\),
  \(\displaystyle {10^{-3}}\),
  \(\displaystyle {10^{-2}}\),
  \(\displaystyle {10^{-1}}\),
  \(\displaystyle {10^{0}}\),
  \(\displaystyle {10^{1}}\)
}
]
\addplot [semithick, blue, dash pattern=on 5.55pt off 2.4pt, mark=o, mark size=2.5, mark repeat=2, mark options={solid,fill opacity=0}]
table {%
-10 0.496585365853659
-9.5 0.497042682926829
-9 0.496768292682927
-8.5 0.498323170731707
-8 0.49625
-7.5 0.494969512195122
-7 0.484359756097561
-6.5 0.480884146341463
-6 0.462256097560976
-5.5 0.443384146341463
-5 0.386219512195122
-4.5 0.295
-4 0.215609756097561
-3.5 0.115396341463415
-3 0.0660975609756098
-2.5 0.0301219512195122
-2 0.0140955284552846
-1.5 0.0112118902439024
-1 0.00641463414634146
-0.5 0.00251641651031895
0 0.00146341463414634
0.5 0.00084087332808812
1 0.000581499107674004
1.5 0.000249518613607189
2 0.000166095020325203
2.5 6.55487804878049e-05
3 3.07926829268293e-05
3.5 0
4 0
4.5 0
5 0
5.5 0
6 0
6.5 0
7 0
7.5 0
8 0
8.5 0
9 0
9.5 0
10 0
10.5 0
11 0
11.5 0
12 0
12.5 0
13 0
13.5 0
14 0
14.5 0
15 0
15.5 0
16 0
16.5 0
17 0
17.5 0
18 0
18.5 0
19 0
19.5 0
20 0
20.5 0
21 0
21.5 0
22 0
22.5 0
23 0
23.5 0
24 0
24.5 0
25 0
25.5 0
26 0
26.5 0
27 0
27.5 0
28 0
28.5 0
29 0
29.5 0
30 0
30.5 0
31 0
31.5 0
32 0
32.5 0
33 0
33.5 0
34 0
34.5 0
35 0
35.5 0
36 0
36.5 0
37 0
37.5 0
38 0
38.5 0
39 0
39.5 0
};
\addlegendentry{NPD with CP}
\addplot [semithick, red, dash pattern=on 5.55pt off 2.4pt, mark=o, mark size=2.5, mark repeat=2, mark options={solid,fill opacity=0}]
table {%
-10 0.496219512195122
-9.5 0.500457317073171
-9 0.492774390243902
-8.5 0.495701219512195
-8 0.497743902439024
-7.5 0.497012195121951
-7 0.484390243902439
-6.5 0.48734756097561
-6 0.481341463414634
-5.5 0.473536585365854
-5 0.417682926829268
-4.5 0.367134146341463
-4 0.273841463414634
-3.5 0.1975
-3 0.0916189024390244
-2.5 0.0571493902439024
-2 0.0161585365853659
-1.5 0.009189735772357724
-1 0.00767073170731707
-0.5 0.00289634146341463
0 0.0023608993902439
0.5 0.00103560188827695
1 0.000788671373555841
1.5 0.000578658536585366
2 0.00031612950058072
2.5 0.000228353658536585
3 2.80487804878049e-05
3.5 0
4 0
4.5 0
5 0
5.5 0
6 0
6.5 0
7 0
7.5 0
8 0
8.5 0
9 0
9.5 0
10 0
10.5 0
11 0
11.5 0
12 0
12.5 0
13 0
13.5 0
14 0
14.5 0
15 0
15.5 0
16 0
16.5 0
17 0
17.5 0
18 0
18.5 0
19 0
19.5 0
20 0
20.5 0
21 0
21.5 0
22 0
22.5 0
23 0
23.5 0
24 0
24.5 0
25 0
25.5 0
26 0
26.5 0
27 0
27.5 0
28 0
28.5 0
29 0
29.5 0
30 0
30.5 0
31 0
31.5 0
32 0
32.5 0
33 0
33.5 0
34 0
34.5 0
35 0
35.5 0
36 0
36.5 0
37 0
37.5 0
38 0
38.5 0
39 0
39.5 0
};
\addlegendentry{NPD w/o CP}
\addplot [semithick, blue, mark=*, mark size=2.5, mark repeat=2, mark options={solid}]
table {%
-10 0.49656862745098
-9.5 0.499052287581699
-9 0.498267973856209
-8.5 0.500490196078431
-8 0.498366013071895
-7.5 0.501045751633987
-7 0.502450980392157
-6.5 0.503954248366013
-6 0.500816993464052
-5.5 0.503006535947712
-5 0.496470588235294
-4.5 0.50202614379085
-4 0.497254901960784
-3.5 0.494705882352941
-3 0.495718954248366
-2.5 0.496732026143791
-2 0.48
-1.5 0.467777777777778
-1 0.444052287581699
-0.5 0.344052287581699
0 0.190980392156863
0.5 0.129607843137255
1 0.0444117647058824
1.5 0.0149074074074074
2 0.00669934640522876
2.5 0.00276573787409701
3 0.000930392156862745
3.5 0.000374509803921569
4 0
4.5 0
5 0
5.5 0
6 0
6.5 0
7 0
7.5 0
8 0
8.5 0
9 0
9.5 0
10 0
10.5 0
11 0
11.5 0
12 0
12.5 0
13 0
13.5 0
14 0
14.5 0
15 0
15.5 0
16 0
16.5 0
17 0
17.5 0
18 0
18.5 0
19 0
19.5 0
20 0
20.5 0
21 0
21.5 0
22 0
22.5 0
23 0
23.5 0
24 0
24.5 0
25 0
25.5 0
26 0
26.5 0
27 0
27.5 0
28 0
28.5 0
29 0
29.5 0
30 0
30.5 0
31 0
31.5 0
32 0
32.5 0
33 0
33.5 0
34 0
34.5 0
35 0
35.5 0
36 0
36.5 0
37 0
37.5 0
38 0
38.5 0
39 0
39.5 0
};
\addlegendentry{5G Polar with CP}
\addplot [semithick, red, mark=*, mark size=2.5, mark repeat=2, mark options={solid}]
table {%
-10 0.500228758169935
-9.5 0.502777777777778
-9 0.501666666666667
-8.5 0.495980392156863
-8 0.499705882352941
-7.5 0.501241830065359
-7 0.502745098039216
-6.5 0.500882352941176
-6 0.501960784313725
-5.5 0.50218954248366
-5 0.502385620915033
-4.5 0.50062091503268
-4 0.504183006535948
-3.5 0.499183006535948
-3 0.502941176470588
-2.5 0.498235294117647
-2 0.490915032679739
-1.5 0.489019607843137
-1 0.489281045751634
-0.5 0.479542483660131
0 0.457058823529412
0.5 0.437712418300654
1 0.360130718954248
1.5 0.252843137254902
2 0.156928104575163
2.5 0.0946078431372549
3 0.0338126361655773
3.5 0.0190522875816993
4 0.0111968954248366
4.5 0.00522712418300654
5 0.00358094519859226
5.5 0.0015582788671024
6 0.000815659597172202
6.5 0.000401307189542484
7 0.000338888888888889
7.5 0
8 0
8.5 0
9 0
9.5 0
10 0
10.5 0
11 0
11.5 0
12 0
12.5 0
13 0
13.5 0
14 0
14.5 0
15 0
15.5 0
16 0
16.5 0
17 0
17.5 0
18 0
18.5 0
19 0
19.5 0
20 0
20.5 0
21 0
21.5 0
22 0
22.5 0
23 0
23.5 0
24 0
24.5 0
25 0
25.5 0
26 0
26.5 0
27 0
27.5 0
28 0
28.5 0
29 0
29.5 0
30 0
30.5 0
31 0
31.5 0
32 0
32.5 0
33 0
33.5 0
34 0
34.5 0
35 0
35.5 0
36 0
36.5 0
37 0
37.5 0
38 0
38.5 0
39 0
39.5 0
};
\addlegendentry{5G Polar w/o CP}

\end{axis}

\end{tikzpicture}}
\caption{BER comparison of NPD and 5G polar decoders with and without CP over a TDL-A channel with 300\,ns delay spread. The OFDM system uses BPSK, \(k = 153\) information bits, and code length \(N = 512\).}
\label{fig:BPSK_TDL-A_cp}
\end{figure}

\subsection{Performance Without Cyclic Prefix}

This experiment evaluates the impact of removing the CP in OFDM systems by comparing the performance of the proposed NPD with that of a 5G polar decoder. To show that, the experiment was conducted over a scenario characterized by a large delay spread of \(300\,\text{ns}\), where the CP is essential for mitigating ISI. Both systems use BPSK modulation, a code length of \( N = 512 \), and \( k = 153 \) information bits.

\textbf{Insight}: This experiment shows that the removal of the CP has a negligible effect on the NPD's performance. The BER curves for the NPD with and without CP are nearly identical, indicating that the NPD can effectively mitigate ISI without relying on a CP. This resilience allows for a reduction in the transmission overhead while maintaining high decoding reliability.

Figure \ref{fig:BPSK_TDL-A_cp} shows the BER performance versus SNR for both the 5G polar decoder and NPD with and without CP. The 5G decoder suffers significant degradation when the CP is removed, highlighting its dependence on the CP for mitigating ISI. In contrast, the NPD maintains a consistently strong performance across the entire SNR range and outperforms the 5G decoder in both configurations. The near overlap of the NPD curves confirms its capability to learn and compensate for the ISI in the absence of a CP.

\begin{figure}[h!]
\centering
\scalebox{.85}{
\begin{tikzpicture}

\definecolor{darkgray176}{RGB}{176,176,176}
\definecolor{green01270}{RGB}{0,127,0}
\definecolor{lightgray204}{RGB}{204,204,204}

\begin{axis}[
legend cell align={left},
legend style={fill opacity=0.8,   font=\footnotesize,
draw opacity=1, text opacity=1, draw=lightgray204},
log basis y={10},
tick align=outside,
tick pos=left,
unbounded coords=jump,
x grid style={darkgray176},
xlabel={SNR [dB]},
xmajorgrids,
xmin=-10, xmax=3,
xtick style={color=black},
y grid style={darkgray176},
ylabel={BLER},
ymajorgrids,
ymin=0.001, ymax=1,
ymode=log,
ytick style={color=black},
ytick={0.0001,0.001,0.01,0.1,1,10},
yticklabels={
  \(\displaystyle {10^{-4}}\),
  \(\displaystyle {10^{-3}}\),
  \(\displaystyle {10^{-2}}\),
  \(\displaystyle {10^{-1}}\),
  \(\displaystyle {10^{0}}\),
  \(\displaystyle {10^{1}}\)
}
]

\addplot [semithick, red, mark=*, mark size=2.5, mark repeat=2, mark options={solid}]
table {%
-10 1
-9.5 1
-9 1
-8.5 1
-8 1
-7.5 1
-7 1
-6.5 1
-6 0.995
-5.5 0.985
-5 0.85
-4.5 0.795
-4 0.485
-3.5 0.295
-3 0.185
-2.5 0.116666666666667
-2 0.0675
-1.5 0.06875
-1 0.0466666666666667
-0.5 0.027
0 0.0255
0.5 0.0288888888888889
1 0.0193333333333333
1.5 0.0208333333333333
2 0.0173333333333333
2.5 0.0178571428571429
3 0.0166666666666667
3.5 0.0119047619047619
4 0.01275
4.5 0.0131578947368421
5 0.0108695652173913
5.5 0.0115909090909091
6 nan
6.5 nan
7 nan
7.5 nan
8 nan
8.5 nan
9 nan
9.5 nan
10 nan
10.5 nan
11 nan
11.5 nan
12 nan
12.5 nan
13 nan
13.5 nan
14 nan
14.5 nan
15 nan
15.5 nan
16 nan
16.5 nan
17 nan
17.5 nan
18 nan
18.5 nan
19 nan
19.5 nan
};
\addlegendentry{5G Polar $1P$}
\addplot [semithick, blue, mark=asterisk, mark size=2.5, mark repeat=2, mark options={solid}]
table {%
-10 1
-9.5 1
-9 1
-8.5 1
-8 1
-7.5 1
-7 1
-6.5 1
-6 1
-5.5 1
-5 0.975
-4.5 0.845
-4 0.615
-3.5 0.25
-3 0.085
-2.5 0.0089285714285714
-2 0.0006925207756232
-1.5 0
-1 0
-0.5 0
0 0
0.5 0
1 0
1.5 0
2 0
2.5 0
3 0
3.5 0
4 0
4.5 0
5 0
5.5 0
6 nan
6.5 nan
7 nan
7.5 nan
8 nan
8.5 nan
9 nan
9.5 nan
10 nan
10.5 nan
11 nan
11.5 nan
12 nan
12.5 nan
13 nan
13.5 nan
14 nan
14.5 nan
15 nan
15.5 nan
16 nan
16.5 nan
17 nan
17.5 nan
18 nan
18.5 nan
19 nan
19.5 nan
};
\addlegendentry{5G Polar $2P$}
\addplot [semithick, green01270, dash pattern=on 5.55pt off 2.4pt, mark=triangle, mark size=2.5, mark repeat=2, mark options={solid,fill opacity=0}]
table {%
-10 0.935
-9.5 0.77
-9 0.62
-8.5 0.45
-8 0.255
-7.5 0.1425
-7 0.0625
-6.5 0.0371428571428571
-6 0.0236363636363636
-5.5 0.0131578947368421
-5 0.0055555555555555
-4.5 0.0028735632183908
-4 0.0013513513513513
-3.5 0.0006265664160401
-3 0
-2.5 0
-2 0
-1.5 0
-1 0
-0.5 0
0 0
0.5 0
1 0
1.5 0
2 0
2.5 0
3 0
3.5 0
4 0
4.5 0
5 0
5.5 0
6 nan
6.5 nan
7 nan
7.5 nan
8 nan
8.5 nan
9 nan
9.5 nan
10 nan
10.5 nan
11 nan
11.5 nan
12 nan
12.5 nan
13 nan
13.5 nan
14 nan
14.5 nan
15 nan
15.5 nan
16 nan
16.5 nan
17 nan
17.5 nan
18 nan
18.5 nan
19 nan
19.5 nan
};
\addlegendentry{NPD}
\end{axis}

\end{tikzpicture}}
\caption{BLER performance under high Doppler conditions. The systems employed BPSK modulation, an OFDM, $k=102$ information bits, and a code length $N=1024$. 5G polar code includes 1P and 2P pilot patterns.}
\label{fig:doppler-A_delay_spreads-bler}
\end{figure}
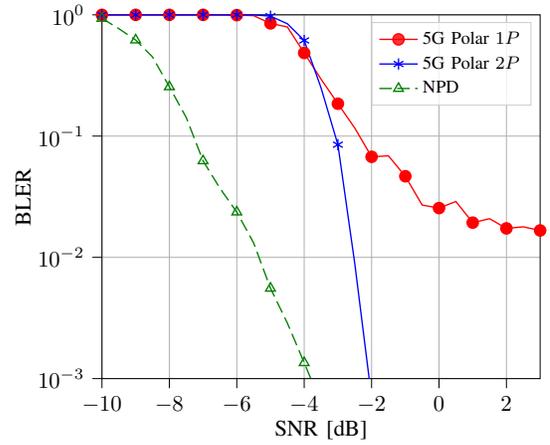

\begin{figure}[h!]
\centering
\scalebox{.85}{
\begin{tikzpicture}

\definecolor{darkgray176}{RGB}{176,176,176}
\definecolor{darkviolet1910191}{RGB}{191,0,191}
\definecolor{green01270}{RGB}{0,127,0}
\definecolor{lightgray204}{RGB}{204,204,204}

\begin{axis}[
legend cell align={left},
legend style={
  font=\footnotesize,
  fill opacity=0.8,
  draw opacity=1,
  text opacity=1,
  at={(0.03,0.03)},
  anchor=south west,
  draw=lightgray204
},
log basis y={10},
tick align=outside,
tick pos=left,
x grid style={darkgray176},
xlabel={SNR [dB]},
xmajorgrids,
xmin=-7, xmax=2,
xtick style={color=black},
y grid style={darkgray176},
ylabel={BER},
ymajorgrids,
ymin=0.001, ymax=0.49,
ymode=log,
ytick style={color=black},
ytick={0.0001,0.001,0.01,0.1,1,10},
yticklabels={
  \(\displaystyle {10^{-4}}\),
  \(\displaystyle {10^{-3}}\),
  \(\displaystyle {10^{-2}}\),
  \(\displaystyle {10^{-1}}\),
  \(\displaystyle {10^{0}}\),
  \(\displaystyle {10^{1}}\)
}
]
\addlegendimage{empty legend}
\addlegendentry{\hspace{-.6cm}\underline{Decoder / Delay}}
\addplot [semithick, green01270, dash pattern=on 5.55pt off 2.4pt, mark=o, mark size=2.5, mark repeat=2, mark options={solid,fill opacity=0}]
table {%
-10 0.4984375
-9.5 0.50154296875
-9 0.4959765625
-8.5 0.48814453125
-8 0.4929296875
-7.5 0.48421875
-7 0.470234375
-6.5 0.4532421875
-6 0.367265625
-5.5 0.29837890625
-5 0.1890234375
-4.5 0.08876953125
-4 0.06001953125
-3.5 0.03033203125
-3 0.01662109375
-2.5 0.0097509765625
-2 0.004578125
-1.5 0.00253689236111111
-1 0.000891927083333333
-0.5 0.000719962284482759
0 0.000293817934782609
0.5 0.000131072998046875
1 3.91273878737542e-05
1.5 0
2 0
2.5 0
3 0
3.5 0
4 0
4.5 0
5 0
5.5 0
6 0
6.5 0
7 0
7.5 0
8 0
8.5 0
9 0
9.5 0
10 0
10.5 0
11 0
11.5 0
12 0
12.5 0
13 0
13.5 0
14 0
14.5 0
15 0
15.5 0
16 0
16.5 0
17 0
17.5 0
18 0
18.5 0
19 0
19.5 0
};
\addlegendentry{NPD 300ns }
\addplot [semithick, blue, dash pattern=on 5.55pt off 2.4pt, mark=o, mark size=2.5, mark repeat=2, mark options={solid,fill opacity=0}]
table {%
-10 0.4993359375
-9.5 0.49849609375
-9 0.4990234375
-8.5 0.49333984375
-8 0.48947265625
-7.5 0.4829296875
-7 0.45599609375
-6.5 0.3953125
-6 0.31962890625
-5.5 0.18443359375
-5 0.09712890625
-4.5 0.04548828125
-4 0.020615234375
-3.5 0.0142643229166667
-3 0.00525948660714286
-2.5 0.0019873046875
-2 0.000830078125
-1.5 0.000546875
-1 0.000151163736979167
-0.5 9.6923828125e-05
0 2.6758672907489e-05
0.5 0
1 0
1.5 0
2 0
2.5 0
3 0
3.5 0
4 0
4.5 0
5 0
5.5 0
6 0
6.5 0
7 0
7.5 0
8 0
8.5 0
9 0
9.5 0
10 0
10.5 0
11 0
11.5 0
12 0
12.5 0
13 0
13.5 0
14 0
14.5 0
15 0
15.5 0
16 0
16.5 0
17 0
17.5 0
18 0
18.5 0
19 0
19.5 0
};
\addlegendentry{NPD 100ns }

\addplot [semithick, red, dash pattern=on 5.55pt off 2.4pt, mark=o, mark size=2.5, mark repeat=2, mark options={solid,fill opacity=0}]
table {%
-10 0.5005078125
-9.5 0.4958203125
-9 0.495078125
-8.5 0.48982421875
-8 0.491953125
-7.5 0.47970703125
-7 0.46818359375
-6.5 0.392890625
-6 0.26939453125
-5.5 0.11080078125
-5 0.0252994791666667
-4.5 0.011396484375
-4 0.00209077380952381
-3.5 0.00119562922297297
-3 0.000720525568181818
-2.5 0.000348524305555556
-2 0.000267145553691275
-1.5 0.000117625420403587
-1 6.97115384615385e-05
-0.5 0
0 0
0.5 0
1 0
1.5 0
2 0
2.5 0
3 0
3.5 0
4 0
4.5 0
5 0
5.5 0
6 0
6.5 0
7 0
7.5 0
8 0
8.5 0
9 0
9.5 0
10 0
10.5 0
11 0
11.5 0
12 0
12.5 0
13 0
13.5 0
14 0
14.5 0
15 0
15.5 0
16 0
16.5 0
17 0
17.5 0
18 0
18.5 0
19 0
19.5 0
};
\addlegendentry{NPD 10ns }
\addplot [semithick, green01270, mark=*, mark size=2.5, mark repeat=2, mark options={solid}]
table {%
-10 0.50140625
-9.5 0.4992578125
-9 0.49962890625
-8.5 0.5023046875
-8 0.50109375
-7.5 0.50001953125
-7 0.49345703125
-6.5 0.5037890625
-6 0.49986328125
-5.5 0.4998828125
-5 0.4989453125
-4.5 0.49439453125
-4 0.4973046875
-3.5 0.501875
-3 0.49466796875
-2.5 0.48634765625
-2 0.47630859375
-1.5 0.4153125
-1 0.2566015625
-0.5 0.10892578125
0 0.029443359375
0.5 0.00701729910714286
1 0.00152242726293103
1.5 0.000341796875
2 0
2.5 0
3 0
3.5 0
4 0
4.5 0
5 0
5.5 0
6 0
6.5 0
7 0
7.5 0
8 0
8.5 0
9 0
9.5 0
10 0
10.5 0
11 0
11.5 0
12 0
12.5 0
13 0
13.5 0
14 0
14.5 0
15 0
15.5 0
16 0
16.5 0
17 0
17.5 0
18 0
18.5 0
19 0
19.5 0
};
\addlegendentry{5G Polar 300ns }
\addplot [semithick, blue, mark=*, mark size=2.5, mark repeat=2, mark options={solid}]
table {%
-10 0.49958984375
-9.5 0.50265625
-9 0.498671875
-8.5 0.50220703125
-8 0.5012890625
-7.5 0.4996484375
-7 0.49978515625
-6.5 0.5007421875
-6 0.49951171875
-5.5 0.4964453125
-5 0.49681640625
-4.5 0.4994921875
-4 0.50015625
-3.5 0.49599609375
-3 0.49107421875
-2.5 0.48427734375
-2 0.4630859375
-1.5 0.38982421875
-1 0.1974609375
-0.5 0.0715625
0 0.014208984375
0.5 0.00248574746621622
1 0.000852072010869565
1.5 0.000200385551948052
2 0
2.5 0
3 0
3.5 0
4 0
4.5 0
5 0
5.5 0
6 0
6.5 0
7 0
7.5 0
8 0
8.5 0
9 0
9.5 0
10 0
10.5 0
11 0
11.5 0
12 0
12.5 0
13 0
13.5 0
14 0
14.5 0
15 0
15.5 0
16 0
16.5 0
17 0
17.5 0
18 0
18.5 0
19 0
19.5 0
};
\addlegendentry{5G Polar 100ns }

\addplot [semithick, red, mark=*, mark size=2.5, mark repeat=2, mark options={solid}]
table {%
-10 0.5018359375
-9.5 0.50095703125
-9 0.50140625
-8.5 0.50283203125
-8 0.4971484375
-7.5 0.49978515625
-7 0.50064453125
-6.5 0.4990625
-6 0.5022265625
-5.5 0.50033203125
-5 0.50234375
-4.5 0.49810546875
-4 0.4966015625
-3.5 0.4959375
-3 0.49435546875
-2.5 0.48388671875
-2 0.4416015625
-1.5 0.3216015625
-1 0.13064453125
-0.5 0.0193994140625
0 0.00124399038461538
0.5 0.000121328125
1 0
1.5 0
2 0
2.5 0
3 0
3.5 0
4 0
4.5 0
5 0
5.5 0
6 0
6.5 0
7 0
7.5 0
8 0
8.5 0
9 0
9.5 0
10 0
10.5 0
11 0
11.5 0
12 0
12.5 0
13 0
13.5 0
14 0
14.5 0
15 0
15.5 0
16 0
16.5 0
17 0
17.5 0
18 0
18.5 0
19 0
19.5 0
};
\addlegendentry{5G Polar 10ns }
\end{axis}

\end{tikzpicture}}
\caption{BER performance across various delay spreads. The system employs OFDM with BPSK modulation, \(k = 256\) information bits, and code length \(N = 1024\).}
\label{fig:BPSK_TDL-A_delay_spreads-ber}
\end{figure}
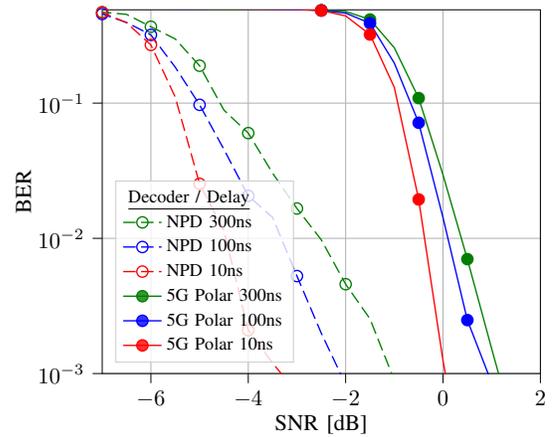

\subsection{Robust Performance Under Varying Channel Conditions}

In this set of experiments, we evaluated the robustness of the proposed NPD under challenging channel conditions and compared it with that of the 5G polar decoder. Experiments were conducted under both high and low Doppler conditions and across multiple delay spreads. Notably, the NPD was trained on TDL-C and evaluated on TDL-A to demonstrate generalization across different propagation configurations. All tests used an OFDM waveform with BPSK modulation.

\textbf{Insight}: These experiments demonstrate that the NPD has strong robustness under diverse channel conditions, including varying SNRs, Doppler effects, and multipath propagation, without requiring retraining, fine-tuning, or the use of pilots or CP. It consistently outperformed the 5G polar decoder across all tested scenarios. These results highlight the practical suitability of the NPD for dynamic wireless environments.

Figure \ref{fig:doppler-A_delay_spreads-bler} presents the BLER performance under high Doppler as a function of SNR. As expected, the 5G decoder with a two-pilot ("$2P$") configuration outperforms its one-pilot ("$1P$") variant owing to its improved resilience to channel aging caused by the Doppler effect. However, the NPD, despite being completely pilotless, achieves a lower BLER across all SNR values, demonstrating strong robustness against time-varying channels. Figure \ref{fig:BPSK_TDL-A_delay_spreads-ber} shows the BER performance for various TDL-A delay spreads as a function of SNR. While both decoders benefit from shorter delay spreads, the NPD consistently outperforms the 5G decoder in all configurations.

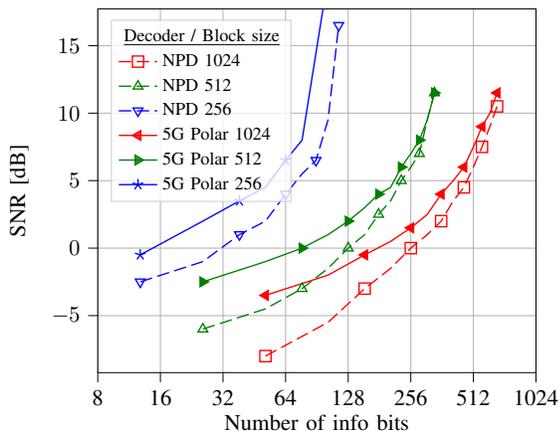
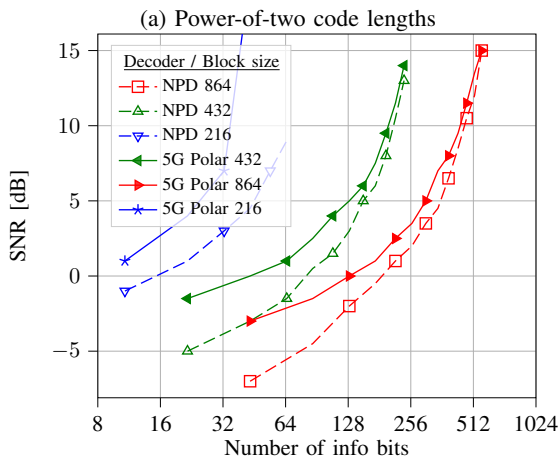
\begin{figure}[h!]
\centering
\subfloat[Power-of-two code lengths]{%
    \scalebox{0.85}{
\begin{tikzpicture}

\definecolor{darkgray176}{RGB}{176,176,176}
\definecolor{green01270}{RGB}{0,127,0}
\definecolor{lightgray204}{RGB}{204,204,204}

\begin{axis}[
legend cell align={left},
legend style={
  font=\footnotesize,
  fill opacity=0.8,
  draw opacity=1,
  text opacity=1,
  at={(0.03,0.97)},
  anchor=north west,
  draw=lightgray204
},
tick align=outside,
tick pos=left,
x grid style={darkgray176},
xlabel={Number of info bits},
xmajorgrids,
xmin=3, xmax=10,
xtick style={color=black},
xtick={3,4,5,6,7,8,9,10},
xticklabels={8,16,32,64,128,256,512,1024},
y grid style={darkgray176},
ylabel={SNR [dB]},
ymajorgrids,
ymin=-9.225, ymax=17.725,
ytick style={color=black}
]
\addlegendimage{empty legend}
\addlegendentry{\hspace{-.6cm}\underline{Decoder /  Block size}}
\addplot [semithick, red, dash pattern=on 5.55pt off 2.4pt, mark=square, mark size=2.5, mark repeat=2, mark options={solid,fill opacity=0}]
table {%
5.67807190511264 -8
6.67807190511264 -5.5
7.26303440583379 -3
7.67807190511264 -1.5
8 0
8.26303440583379 1
8.48542682717024 2
8.67807190511264 3.5
8.84799690655495 4.5
9 6
9.13750352374993 7.5
9.26303440583379 9
9.37851162325373 10.5
};
\addlegendentry{NPD 1024 }
\addplot [semithick, green01270, dash pattern=on 5.55pt off 2.4pt, mark=triangle, mark size=2.5, mark repeat=2, mark options={solid,fill opacity=0}]
table {%
4.67807190511264 -6
5.67807190511264 -4.5
6.26303440583379 -3
6.67807190511264 -1.5
7 0
7.26303440583379 1
7.48542682717024 2.5
7.67807190511264 3.5
7.84799690655495 5
8 6
8.13750352374993 7
8.26303440583379 9.5
8.37851162325373 11.5
};
\addlegendentry{NPD 512 }
\addplot [semithick, blue, dash pattern=on 5.55pt off 2.4pt, mark=triangle, mark size=2.5, mark repeat=2, mark options={solid,rotate=180,fill opacity=0}]
table {%
3.67807190511264 -2.5
4.67807190511264 -1
5.26303440583379 1
5.67807190511264 2
6 4
6.26303440583379 5.5
6.48542682717024 6.5
6.67807190511264 9.5
6.84799690655495 16.5
};
\addlegendentry{NPD 256 }
\addplot [semithick, red, mark=triangle*, mark size=2.5, mark repeat=2, mark options={solid,rotate=90}]
table {%
5.67807190511264 -3.5
6.67807190511264 -2
7.26303440583379 -0.5
7.67807190511264 0.5
8 1.5
8.26303440583379 2.5
8.48542682717024 4
8.67807190511264 5
8.84799690655495 6
9 7.5
9.13750352374993 9
9.26303440583379 10
9.37851162325373 11.5
};
\addlegendentry{5G Polar 1024 }
\addplot [semithick, green01270, mark=triangle*, mark size=2.5, mark repeat=2, mark options={solid,rotate=270}]
table {%
4.67807190511264 -2.5
5.67807190511264 -1
6.26303440583379 0
6.67807190511264 1
7 2
7.26303440583379 3
7.48542682717024 4
7.67807190511264 4.5
7.84799690655495 6
8 7
8.13750352374993 8
8.26303440583379 9.5
8.37851162325373 11.5
};
\addlegendentry{5G Polar 512 }
\addplot [semithick, blue, mark=star, mark size=2.5, mark repeat=2, mark options={solid}]
table {%
3.67807190511264 -0.5
4.67807190511264 2
5.26303440583379 3.5
5.67807190511264 4.5
6 6.5
6.26303440583379 8
6.84799690655495 25
};
\addlegendentry{5G Polar 256 }
\end{axis}

\end{tikzpicture}}%
    \label{fig:power-of-two}
}
\hfill
\subfloat[Non-power-of-two code lengths with rate matching]{%
    \scalebox{0.85}{
\begin{tikzpicture}

\definecolor{darkgray176}{RGB}{176,176,176}
\definecolor{green01270}{RGB}{0,127,0}
\definecolor{lightgray204}{RGB}{204,204,204}

\begin{axis}[
legend cell align={left},
legend style={
  font=\footnotesize,
  fill opacity=0.8,
  draw opacity=1,
  text opacity=1,
  at={(0.03,0.97)},
  anchor=north west,
  draw=lightgray204
},
tick align=outside,
tick pos=left,
unbounded coords=jump,
x grid style={darkgray176},
xlabel={Number of info bits},
xmajorgrids,
xmin=3, xmax=10,
xtick style={color=black},
xtick={3,4,5,6,7,8,9,10},
xticklabels={8,16,32,64,128,256,512,1024},
y grid style={darkgray176},
ylabel={SNR [dB]},
ymajorgrids,
ymin=-8.1, ymax=16.1,
ytick style={color=black}
]
\addlegendimage{empty legend}
\addlegendentry{\hspace{-.6cm}\underline{Decoder / Block size}}
\addplot [semithick, red, dash pattern=on 5.55pt off 2.4pt, mark=square, mark size=2.5, mark repeat=2, mark options={solid,fill opacity=0}]
table {%
5.43295940727611 -7
6.43295940727611 -4.5
7.01792190799726 -2
7.43295940727611 -0.5
7.75488750216347 1
8.01792190799726 2
8.24031432933371 3.5
8.43295940727611 4.5
8.60288440871842 6.5
8.75488750216347 8.5
8.8923910259134 10.5
9.01792190799726 12
9.1333991254172 15
};
\addlegendentry{NPD 864 }
\addplot [semithick, green01270, dash pattern=on 5.55pt off 2.4pt, mark=triangle, mark size=2.5, mark repeat=2, mark options={solid,fill opacity=0}]
table {%
4.43295940727611 -5
5.43295940727611 -3
6.01792190799726 -1.5
6.43295940727611 0.5
6.75488750216347 1.5
7.01792190799726 3
7.24031432933371 5
7.43295940727611 6
7.60288440871842 8
7.75488750216347 10.5
7.8923910259134 13
8.01792190799726 nan
};
\addlegendentry{NPD 432 }
\addplot [semithick, blue, dash pattern=on 5.55pt off 2.4pt, mark=triangle, mark size=2.5, mark repeat=2, mark options={solid,rotate=180,fill opacity=0}]
table {%
3.43295940727611 -1
4.43295940727611 1
5.01792190799726 3
5.43295940727611 4.5
5.75488750216347 7
6.01792190799726 9
};
\addlegendentry{NPD 216 }
\addplot [semithick, green01270, mark=triangle*, mark size=2.5, mark repeat=2, mark options={solid,rotate=90}]
table {%
4.43295940727611 -1.5
5.43295940727611 0
6.01792190799726 1
6.43295940727611 2.5
6.75488750216347 4
7.01792190799726 5
7.24031432933371 6
7.43295940727611 7.5
7.60288440871842 9.5
7.75488750216347 11.5
7.8923910259134 14
};
\addlegendentry{5G Polar 432 }
\addplot [semithick, red, mark=triangle*, mark size=2.5, mark repeat=2, mark options={solid,rotate=270}]
table {%
5.43295940727611 -3
6.43295940727611 -1.5
7.01792190799726 0
7.43295940727611 1
7.75488750216347 2.5
8.01792190799726 3.5
8.24031432933371 5
8.43295940727611 7
8.60288440871842 8
8.75488750216347 9.5
8.8923910259134 11.5
9.01792190799726 13.5
9.1333991254172 15
};
\addlegendentry{5G Polar 864 }
\addplot [semithick, blue, mark=star, mark size=2.5, mark repeat=2, mark options={solid}]
table {%
3.43295940727611 1
4.43295940727611 4
5.01792190799726 7
5.43295940727611 20

};
\addlegendentry{5G Polar 216 }
\end{axis}

\end{tikzpicture}}%
    \label{fig:non-power-of-two}
}
\caption{SNR required to achieve $\text{BLER} = 10^{-3}$ under BPSK modulation for (a) power-of-two code lengths and (b) non-power-of-two code lengths using rate matching. Both systems use OFDM over a TDL-A channel with 100\,ns delay spread and user mobility between 0–8\,m/s.}
\label{fig:block-size-comparison}
\end{figure}

\subsection{Performance Across Different Code Length}

This experiment assessed the performance of the proposed NPD compared to that of the 5G polar decoder across a range of code lengths, including scenarios involving rate matching for non-power-of-two code lengths and for modulation BPSK and QPSK. To evaluate this, experiments were conducted over a channel with user mobility ranging from 0 to 8 m/s and a nominal delay spread of 100 ns. The 5G decoder applies rate matching as specified by the 3GPP standard, while the NPD utilizes the rate matching approach described in Section~\ref{sec:rm}.

\begin{figure}[h!]
\centering
\scalebox{0.85}{
\begin{tikzpicture}

\definecolor{darkgray176}{RGB}{176,176,176}
\definecolor{green01270}{RGB}{0,127,0}
\definecolor{lightgray204}{RGB}{204,204,204}
\definecolor{darkviolet1910191}{RGB}{191,0,191}

\begin{axis}[
legend cell align={left},
legend style={
  fill opacity=0.8,
  font=\footnotesize,
  draw opacity=1,
  text opacity=1,
  at={(0.02,0.98)},
  anchor=north west,
  draw=lightgray204
},
tick align=outside,
tick pos=left,
x grid style={darkgray176},
xlabel={Number of info bits},
xmajorgrids,
xmin=3, xmax=10,
xtick style={color=black},
xtick={3,4,5,6,7,8,9,10},
xticklabels={8,16,32,64,128,256,512,1024},
y grid style={darkgray176},
ylabel={SNR [dB]},
ymajorgrids,
ymin=-5.5, ymax=16.5,
ytick style={color=black}
]
\addlegendimage{empty legend}
\addlegendentry{\hspace{-.6cm}\underline{Decoder / Block size}}
\addplot [semithick, darkviolet1910191, dash pattern=on 5.55pt off 2.4pt, mark=triangle, mark size=2.5, mark repeat=2, mark options={solid,fill opacity=0}]
table {%
4.67807190511264 0.5
5.67807190511264 5
6.26303440583379 11.5
6.67807190511264 25
};
\addlegendentry{NPD 436}

\addplot [semithick, darkviolet1910191, mark=triangle*, mark size=2.5, mark repeat=2, mark options={solid,rotate=90}]
table {%
4.67807190511264 4.5
5.67807190511264 25
};
\addlegendentry{5G Polar 436}

\addplot [semithick, green01270, dash pattern=on 5.55pt off 2.4pt, mark=triangle, mark size=2.5, mark repeat=2, mark options={solid,fill opacity=0}]
table {%
4.67807190511264 -1.5
5.67807190511264 1.5
6.26303440583379 4.5
6.67807190511264 6.5
7 10
7.26303440583379 15.5
};
\addlegendentry{NPD 512}

\addplot [semithick, green01270, mark=triangle*, mark size=2.5, mark repeat=2, mark options={solid,rotate=90}]
table {%
4.67807190511264 2
5.67807190511264 5
6.26303440583379 25
};
\addlegendentry{5G Polar 512}

\addplot [semithick, blue, dash pattern=on 5.55pt off 2.4pt, mark=square, mark size=2.5, mark repeat=2, mark options={solid,fill opacity=0}]
table {%
5.6724253419715 -3
6.6724253419715 0
7.25738784269265 2
7.6724253419715 4
8 6
8.26209484537018 7
8.48381577726426 9.5
8.67595703294175 11.5
8.84549005094438 14
9 19.5
};
\addlegendentry{NPD 864}

\addplot [semithick, blue, mark=triangle*, mark size=2.5, mark repeat=2, mark options={solid,rotate=180}]
table {%
5.6724253419715 0
6.6724253419715 2
7.25738784269265 3.5
7.6724253419715 5.5
8 7
8.26209484537018 8.5
8.48381577726426 10.5
8.67595703294175 20

};
\addlegendentry{5G Polar 864}

\addplot [semithick, red, dash pattern=on 5.55pt off 2.4pt, mark=square, mark size=2.5, mark repeat=2, mark options={solid,fill opacity=0}]
table {%
5.67807190511264 -4.5
6.67807190511264 -1.5
7.26303440583379 0
7.67807190511264 1.5
8 3
8.26303440583379 4
8.48542682717024 5.5
8.67807190511264 6.5
8.84799690655495 8
9 9
9.13750352374993 11
9.26303440583379 12.5
9.37851162325373 14.5
};
\addlegendentry{NPD 1024}

\addplot [semithick, red, mark=triangle*, mark size=2.5, mark repeat=2, mark options={solid,rotate=180}]
table {%
5.67807190511264 -1
6.67807190511264 0.5
7.26303440583379 2
7.67807190511264 3
8 4.5
8.26303440583379 5.5
8.48542682717024 6.5
8.67807190511264 8
8.84799690655495 9
9 10.5
9.13750352374993 12
9.26303440583379 14
};
\addlegendentry{5G Polar 1024}

\end{axis}

\end{tikzpicture}}
\caption{Required SNR to achieve BLER \(=10^{-3}\) for various code sizes under QPSK modulation. The system employs OFDM over a TDL-A channel with 100\,ns delay spread and user mobility between 0–8\,m/s.}
\label{fig:qpsk-block-size}
\end{figure}
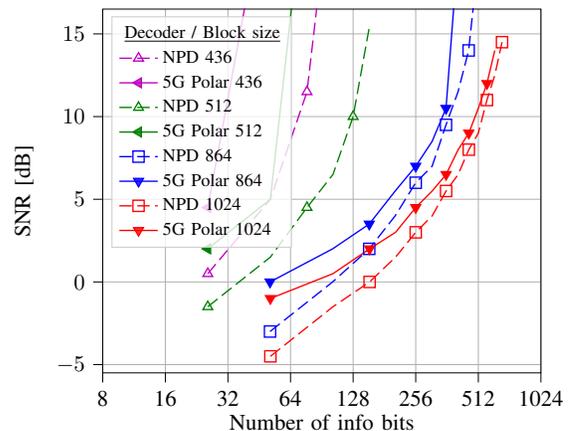

\begin{figure*}[h!]
\centering
\begin{subfigure}[b]{0.48\textwidth}
    \centering
    \scalebox{.85}{
\begin{tikzpicture}

\definecolor{crimson2143940}{RGB}{214,39,40}
\definecolor{darkgray176}{RGB}{176,176,176}
\definecolor{darkorange25512714}{RGB}{255,127,14}
\definecolor{forestgreen4416044}{RGB}{44,160,44}
\definecolor{lightgray204}{RGB}{204,204,204}
\definecolor{mediumpurple148103189}{RGB}{148,103,189}
\definecolor{sienna1408675}{RGB}{140,86,75}
\definecolor{steelblue31119180}{RGB}{31,119,180}
\definecolor{darkgray176}{RGB}{176,176,176}
\definecolor{green01270}{RGB}{0,127,0}
\definecolor{lightgray204}{RGB}{204,204,204}

\begin{axis}[
legend cell align={left},
legend style={
  font=\footnotesize,
  fill opacity=0.8,
  draw opacity=1,
  text opacity=1,
  at={(0.97,0.03)},
  anchor=south east,
  draw=lightgray204
},
tick align=outside,
tick pos=left,
x grid style={darkgray176},
xlabel={SNR [dB]},
xmajorgrids,
xmin=-11.225, xmax=10,
xtick style={color=black},
y grid style={darkgray176},
ylabel={Throughput [Mbit/s]},
ymajorgrids,
ymin=-0.024, ymax=0.504,
ytick style={color=black}
]

\addlegendimage{empty legend}
\addlegendentry{\hspace{-.6cm}\underline{Decoder / REs}}
\addplot [semithick, blue, dash pattern=on 5.55pt off 2.4pt, mark=o, mark size=2.5, mark repeat=4, mark options={solid,fill opacity=0}]
table {%
-10 0
-9.5 0
-9 0
-8.5 0.0024
-8 0.0024
-7.5 0.0048
-7 0.0144
-6.5 0.0168
-6 0.024
-5.5 0.0552
-5 0.0888
-4.5 0.1368
-4 0.1944
-3.5 0.2376
-3 0.2832
-2.5 0.3432
-2 0.3852
-1.5 0.4092
-1 0.432
-0.5 0.4464
0 0.458
0.5 0.4638
1 0.473431578947368
1.5 0.476363636363636
2 0.477894736842105
2.5 0.478461538461538
3 0.479354838709677
3.5 0.4796928
4 0.4797696
4.5 0.4798368
5 0.4798752
5.5 0.4799424
6 0.48
6.5 0.48
7 0.48
7.5 0.48
8 0.48
8.5 0.48
9 0.48
9.5 0.48
10 0.48
10.5 0.48
11 0.48
11.5 0.48
12 0.48
12.5 0.48
13 0.48
13.5 0.48
14 0.48
14.5 0.48
};
\addlegendentry{NPD 192}
\addplot [semithick, green01270, dash pattern=on 5.55pt off 2.4pt, mark=square, mark size=2.5, mark repeat=4, mark options={solid,fill opacity=0}]
table {%
-10 0
-9.5 0.0024
-9 0.0192
-8.5 0.0288
-8 0.0576
-7.5 0.0888
-7 0.1224
-6.5 0.204
-6 0.2352
-5.5 0.264
-5 0.3672
-4.5 0.4188
-4 0.4288
-3.5 0.45216
-3 0.4564
-2.5 0.468
-2 0.47136
-1.5 0.476129032258065
-1 0.478571428571429
-0.5 0.479154929577465
0 0.4795296
0.5 0.4797216
1 0.4798464
1.5 0.4799712
2 0.48
2.5 0.48
3 0.48
3.5 0.48
4 0.48
4.5 0.48
5 0.48
5.5 0.48
6 0.48
6.5 0.48
7 0.48
7.5 0.48
8 0.48
8.5 0.48
9 0.48
9.5 0.48
10 0.48
10.5 0.48
11 0.48
11.5 0.48
12 0.48
12.5 0.48
13 0.48
13.5 0.48
14 0.48
14.5 0.48
};
\addlegendentry{NPD 256}
\addplot [semithick, red, dash pattern=on 5.55pt off 2.4pt, mark=diamond, mark size=2.5, mark repeat=4, mark options={solid,fill opacity=0}]
table {%
-10 0.0792
-9.5 0.1056
-9 0.1584
-8.5 0.276
-8 0.2856
-7.5 0.3504
-7 0.3816
-6.5 0.4128
-6 0.4494
-5.5 0.462514285714286
-5 0.46728
-4.5 0.474
-4 0.4764
-3.5 0.478725
-3 0.4797312
-2.5 0.4799424
-2 0.48
-1.5 0.48
-1 0.48
-0.5 0.48
0 0.48
0.5 0.48
1 0.48
1.5 0.48
2 0.48
2.5 0.48
3 0.48
3.5 0.48
4 0.48
4.5 0.48
5 0.48
5.5 0.48
6 0.48
6.5 0.48
7 0.48
7.5 0.48
8 0.48
8.5 0.48
9 0.48
9.5 0.48
10 0.48
10.5 0.48
11 0.48
11.5 0.48
12 0.48
12.5 0.48
13 0.48
13.5 0.48
14 0.48
14.5 0.48
};
\addlegendentry{NPD 384}
\addplot [semithick, blue, mark=*, mark size=2.5, mark repeat=4, mark options={solid}]
table {%
-10 0
-9.5 0
-9 0
-8.5 0
-8 0
-7.5 0
-7 0
-6.5 0
-6 0
-5.5 0
-5 0
-4.5 0
-4 0.00224225720558696
-3.5 0.00672677161676088
-3 0.00896902882234784
-2.5 0.0134535432335218
-2 0.0269070864670435
-1.5 0.0403606297005652
-1 0.114355117484935
-0.5 0.114355117484935
0 0.217498948941935
0.5 0.215256691736348
1 0.2578595786425
1.5 0.295977951137478
2 0.354276638482739
2.5 0.3610034100995
3 0.392395010977718
3.5 0.411080487690942
4 0.423338160414818
4.5 0.427897416732844
5 0.436567477927781
5.5 0.438174428925118
6 0.442432750723448
6.5 0.443355402013785
7 0.443877236417994
7.5 0.444585480418104
8 0.446209183911805
8.5 0.446295424573558
9 0.445501102688988
9.5 0.446066061111448
10 0.446802722583872
10.5 0.447219431663772
11 0.446582893446069
11.5 0.447039649543503
12 0.446802722583872
12.5 0.447166552156887
13 0.447245926490732
13.5 0.447441415349109
14 0.447147803207167
14.5 0.447295638434099
};
\addlegendentry{5G-Polar-$2P_2$ 192}
\addplot [semithick, green01270, mark=square*, mark size=2.5, mark repeat=4, mark options={solid}]
table {%
-10 0
-9.5 0
-9 0
-8.5 0
-8 0
-7.5 0
-7 0
-6.5 0
-6 0
-5.5 0.00224225720558696
-5 0.00672677161676088
-4.5 0.0134535432335218
-4 0.0336338580838043
-3.5 0.0515719157285
-3 0.0672677161676087
-2.5 0.0986593170458261
-2 0.192834119680478
-1.5 0.215256691736348
-1 0.313916008782174
-0.5 0.353155509879946
0 0.403606297005652
0.5 0.420423226047554
1 0.43785167978189
1.5 0.442098379034895
2 0.445910216284393
2.5 0.447208450709947
3 0.447993837606047
3.5 0.448209277339188
4 0.448361750829168
4.5 0.448397626944457
5 0.448451441117391
5.5 0.448451441117391
6 0.448451441117391
6.5 0.448451441117391
7 0.448451441117391
7.5 0.448451441117391
8 0.448451441117391
8.5 0.448451441117391
9 0.448451441117391
9.5 0.448451441117391
10 0.448451441117391
10.5 0.448451441117391
11 0.448451441117391
11.5 0.448451441117391
12 0.448451441117391
12.5 0.448451441117391
13 0.448451441117391
13.5 0.448451441117391
14 0.448451441117391
14.5 0.448451441117391
};
\addlegendentry{5G-Polar-$2P_2$ 256}
\addplot [semithick, red, mark=diamond*, mark size=2.5, mark repeat=4, mark options={solid}]
table {%
-10 0
-9.5 0
-9 0
-8.5 0
-8 0
-7.5 0.00224225720558696
-7 0.00224225720558696
-6.5 0.00896902882234784
-6 0.00448451441117392
-5.5 0.0246648292614566
-5 0.0515719157285
-4.5 0.114355117484935
-4 0.170411547624609
-3.5 0.22871023496987
-3 0.293735693931891
-2.5 0.363245667305087
-2 0.391273882374924
-1.5 0.430513383472696
-1 0.440883823048536
-0.5 0.445648619610408
0 0.447678248977534
0.5 0.44821824636801
1 0.448415565002102
1.5 0.448451441117391
2 0.448451441117391
2.5 0.448451441117391
3 0.448451441117391
3.5 0.448451441117391
4 0.448451441117391
4.5 0.448451441117391
5 0.448451441117391
5.5 0.448451441117391
6 0.448451441117391
6.5 0.448451441117391
7 0.448451441117391
7.5 0.448451441117391
8 0.448451441117391
8.5 0.448451441117391
9 0.448451441117391
9.5 0.448451441117391
10 0.448451441117391
10.5 0.448451441117391
11 0.448451441117391
11.5 0.448451441117391
12 0.448451441117391
12.5 0.448451441117391
13 0.448451441117391
13.5 0.448451441117391
14 0.448451441117391
14.5 0.448451441117391
};
\addlegendentry{5G Polar $2P_2$ 384}
\end{axis}

\end{tikzpicture}}
    \caption{High Doppler}
    \label{fig:thdophigh}
\end{subfigure}
\hfill
\begin{subfigure}[b]{0.48\textwidth}
    \centering
    \scalebox{.85}{
\begin{tikzpicture}

\definecolor{crimson2143940}{RGB}{214,39,40}
\definecolor{darkgray176}{RGB}{176,176,176}
\definecolor{darkorange25512714}{RGB}{255,127,14}
\definecolor{forestgreen4416044}{RGB}{44,160,44}
\definecolor{lightgray204}{RGB}{204,204,204}
\definecolor{mediumpurple148103189}{RGB}{148,103,189}
\definecolor{sienna1408675}{RGB}{140,86,75}
\definecolor{steelblue31119180}{RGB}{31,119,180}
\definecolor{darkgray176}{RGB}{176,176,176}
\definecolor{green01270}{RGB}{0,127,0}
\definecolor{lightgray204}{RGB}{204,204,204}

\begin{axis}[
legend cell align={left},
legend style={
  font=\footnotesize,
  fill opacity=0.8,
  draw opacity=1,
  text opacity=1,
  at={(0.97,0.03)},
  anchor=south east,
  draw=lightgray204
},
tick align=outside,
tick pos=left,
x grid style={darkgray176},
xlabel={SNR [dB]},
xmajorgrids,
xmin=-11.225, xmax=10,
xtick style={color=black},
y grid style={darkgray176},
ylabel={Throughput [Mbit/s]},
ymajorgrids,
ymin=-0.024, ymax=0.504,
ytick style={color=black}
]

\addlegendimage{empty legend}
\addlegendentry{\hspace{-.6cm}\underline{Decoder / REs}}
\addplot [semithick, blue, dash pattern=on 5.55pt off 2.4pt, mark=o, mark size=2.5, mark repeat=4, mark options={solid,fill opacity=0}]
table {%
-10 0
-9.5 0
-9 0.0024
-8.5 0
-8 0
-7.5 0
-7 0.0144
-6.5 0.024
-6 0.0432
-5.5 0.0672
-5 0.0936
-4.5 0.1296
-4 0.1968
-3.5 0.2712
-3 0.3192
-2.5 0.3936
-2 0.4304
-1.5 0.45
-1 0.466666666666667
-0.5 0.4704
0 0.473557894736842
0.5 0.475466666666667
1 0.477735849056604
1.5 0.47792
2 0.478488888888889
2.5 0.47921568627451
3 0.479154929577465
3.5 0.4796256
4 0.4797312
4.5 0.4798176
5 0.479952
5.5 0.479952
6 0.4799616
6.5 0.48
7 0.48
7.5 0.48
8 0.48
8.5 0.48
9 0.48
9.5 0.48
10 0.48
10.5 0.48
11 0.48
11.5 0.48
12 0.48
12.5 0.48
13 0.48
13.5 0.48
14 0.48
14.5 0.48
};
\addlegendentry{NPD 192}
\addplot [semithick, green01270, dash pattern=on 5.55pt off 2.4pt, mark=square, mark size=2.5, mark repeat=4, mark options={solid,fill opacity=0}]
table {%
-10 0
-9.5 0.0048
-9 0.0192
-8.5 0.0384
-8 0.06
-7.5 0.1056
-7 0.1512
-6.5 0.2184
-6 0.3072
-5.5 0.3456
-5 0.4032
-4.5 0.4482
-4 0.4596
-3.5 0.46752
-3 0.472517647058823
-2.5 0.474
-2 0.47625
-1.5 0.478095238095238
-1 0.479254658385093
-0.5 0.479497907949791
0 0.4798176
0.5 0.4798944
1 0.479952
1.5 0.479856
2 0.4799808
2.5 0.48
3 0.48
3.5 0.48
4 0.48
4.5 0.48
5 0.48
5.5 0.48
6 0.48
6.5 0.48
7 0.48
7.5 0.48
8 0.48
8.5 0.48
9 0.48
9.5 0.48
10 0.48
10.5 0.48
11 0.48
11.5 0.48
12 0.48
12.5 0.48
13 0.48
13.5 0.48
14 0.48
14.5 0.48
};
\addlegendentry{NPD 256}
\addplot [semithick, red, dash pattern=on 5.55pt off 2.4pt, mark=diamond, mark size=2.5, mark repeat=4, mark options={solid,fill opacity=0}]
table {%
-10 0.0672
-9.5 0.1536
-9 0.2208
-8.5 0.2976
-8 0.3504
-7.5 0.3684
-7 0.4176
-6.5 0.4434
-6 0.462857142857143
-5.5 0.470769230769231
-5 0.473684210526316
-4.5 0.478775510204082
-4 0.4796352
-3.5 0.4796928
-3 0.479856
-2.5 0.4798944
-2 0.4799712
-1.5 0.48
-1 0.48
-0.5 0.48
0 0.48
0.5 0.48
1 0.48
1.5 0.48
2 0.48
2.5 0.48
3 0.48
3.5 0.48
4 0.48
4.5 0.48
5 0.48
5.5 0.48
6 0.48
6.5 0.48
7 0.48
7.5 0.48
8 0.48
8.5 0.48
9 0.48
9.5 0.48
10 0.48
10.5 0.48
11 0.48
11.5 0.48
12 0.48
12.5 0.48
13 0.48
13.5 0.48
14 0.48
14.5 0.48
};
\addlegendentry{NPD 384}
\addplot [semithick, blue, mark=*, mark size=2.5, mark repeat=4, mark options={solid}]
table {%
-10 0
-9.5 0
-9 0
-8.5 0
-8 0
-7.5 0
-7 0.00224225720558696
-6.5 0
-6 0
-5.5 0
-5 0.00672677161676088
-4.5 0.00896902882234784
-4 0.0269070864670435
-3.5 0.0358761152893913
-3 0.0560564301396739
-2.5 0.0896902882234783
-2 0.130050917924044
-1.5 0.186107348063717
-1 0.221983463353109
-0.5 0.28028215069837
0 0.320642780398935
0.5 0.368851310319055
1 0.373335824730228
1.5 0.40510113514271
2 0.429018545335638
2.5 0.433035922828981
3 0.439962895981955
3.5 0.441023964123885
4 0.444877843695987
4.5 0.445810560408589
5 0.446915648510825
5.5 0.447247703038603
6 0.447708971844018
6.5 0.447899161017493
7 0.447912436981433
7.5 0.447879436728211
8 0.447870545468276
8.5 0.447993837606047
9 0.448209277339188
9.5 0.448074741906853
10 0.448047834820386
10.5 0.448137525108609
11 0.448092679964498
11.5 0.448128556079787
12 0.448164432195076
12.5 0.448263091512122
13 0.448119587050965
13.5 0.448092679964498
14 0.448173401223899
14.5 0.448164432195076
};
\addlegendentry{5G-Polar-$1P_2$ 192}
\addplot [semithick, green01270, mark=square*, mark size=2.5, mark repeat=4, mark options={solid}]
table {%
-10 0
-9.5 0
-9 0
-8.5 0
-8 0
-7.5 0
-7 0
-6.5 0.00224225720558696
-6 0.00672677161676088
-5.5 0.0179380576446957
-5 0.0381183724949783
-4.5 0.0358761152893913
-4 0.0739944877843696
-3.5 0.0807212594011305
-3 0.1547157471855
-2.5 0.190591862474891
-2 0.255617321436913
-1.5 0.289251179520718
-1 0.355397767085533
-0.5 0.391273882374924
0 0.406969682814033
0.5 0.425580417620404
1 0.436119026486663
1.5 0.440283218439896
2 0.444714345774747
2.5 0.445844165296942
3 0.447351872680036
3.5 0.447791953703984
4 0.448083710935675
4.5 0.44821824636801
5 0.448191339281543
5.5 0.4482541224833
6 0.448281029569767
6.5 0.448352781800346
7 0.448316905685056
7.5 0.448343812771523
8 0.448343812771523
8.5 0.448388657915635
9 0.448451441117391
9.5 0.448451441117391
10 0.448451441117391
10.5 0.448451441117391
11 0.448451441117391
11.5 0.448451441117391
12 0.448451441117391
12.5 0.448451441117391
13 0.448451441117391
13.5 0.448451441117391
14 0.448451441117391
14.5 0.448451441117391
};
\addlegendentry{5G-Polar-$1P_2$ 256}
\addplot [semithick, red, mark=diamond*, mark size=2.5, mark repeat=4, mark options={solid}]
table {%
-10 0.00224225720558696
-9.5 0.00224225720558696
-9 0
-8.5 0.00224225720558696
-8 0.00224225720558696
-7.5 0.00672677161676088
-7 0.0246648292614566
-6.5 0.0448451441117391
-6 0.0717522305787827
-5.5 0.118839631896109
-5 0.156958004391087
-4.5 0.201803148502826
-4 0.266828607464848
-3.5 0.304946979959826
-3 0.371093567524641
-2.5 0.404353716074181
-2 0.425131966179287
-1.5 0.435745316952399
-1 0.444216066395727
-0.5 0.446936402464968
0 0.447803389901904
0.5 0.448146494137432
1 0.448289998598589
1.5 0.448379688886813
2 0.448451441117391
2.5 0.448451441117391
3 0.448451441117391
3.5 0.448451441117391
4 0.448451441117391
4.5 0.448451441117391
5 0.448451441117391
5.5 0.448451441117391
6 0.448451441117391
6.5 0.448451441117391
7 0.448451441117391
7.5 0.448451441117391
8 0.448451441117391
8.5 0.448451441117391
9 0.448451441117391
9.5 0.448451441117391
10 0.448451441117391
10.5 0.448451441117391
11 0.448451441117391
11.5 0.448451441117391
12 0.448451441117391
12.5 0.448451441117391
13 0.448451441117391
13.5 0.448451441117391
14 0.448451441117391
14.5 0.448451441117391
};
\addlegendentry{5G Polar $1P_2$ 384}
\end{axis}

\end{tikzpicture}}
    \caption{Low Doppler}
    \label{fig:thdoplolw}
\end{subfigure}
\caption{Throughput vs. SNR for transmitted information bits $k = 32$ using different REs per frame.}
\label{fig:throughput_doppler}
\end{figure*}
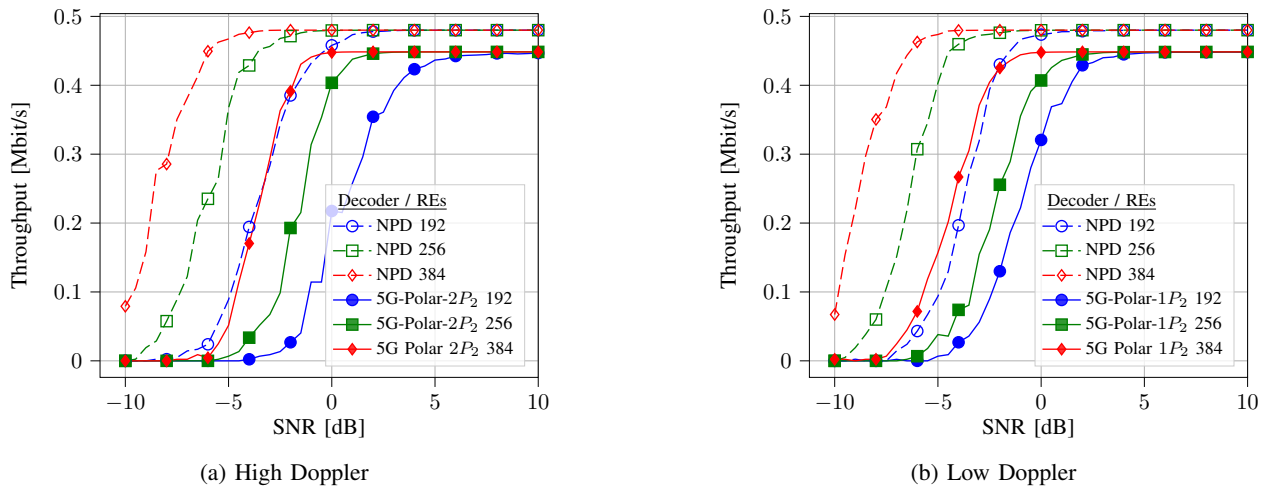

\textbf{Insight}: The NPD achieves superior performance compared to the 5G polar decoder across all evaluated code lengths, including scenarios involving rate matching and QPSK and BPSK modulation schemes. This advantage is particularly evident at low code rates. Notably, the NPD demonstrates strong generalization despite being trained solely on data with a fixed code length of \( N = 1024 \). Its ability to maintain high performance over a wide range of code sizes underscores the robustness of the proposed approach.

Figure \ref{fig:power-of-two} shows the required SNR to achieve a BLER of $10^{-3}$ as a function of the number of information bits for various code sizes, where the code length $N$ is a power of two. Across all configurations, the NPD achieves significantly better performance than the 5G decoder, with a widening SNR gap as the number of information bits is decreased. 

We further evaluated cases in which the code lengths were not powers of two, as shown in Figure \ref{fig:non-power-of-two}. In these scenarios, both decoders employ rate-matching. The NPD continues to deliver superior performance, outperforming the 5G baseline. Finally, Figure \ref{fig:qpsk-block-size} presents the results for QPSK modulation under similar conditions, showing that the NPD maintains its performance advantage over the 5G decoder with and without rate matching.

\begin{figure}[h!]
\centering
\scalebox{.85}{
\begin{tikzpicture}

\definecolor{darkgray176}{RGB}{176,176,176}
\definecolor{darkviolet1910191}{RGB}{191,0,191}
\definecolor{green01270}{RGB}{0,127,0}
\definecolor{lightgray204}{RGB}{204,204,204}

\begin{axis}[
legend cell align={left},
legend style={
    font=\footnotesize,
  fill opacity=0.8,
  draw opacity=1,
  text opacity=1,
  at={(0.03,0.97)},
  anchor=north west,
  draw=lightgray204
},
tick align=outside,
tick pos=left,
x grid style={darkgray176},
xlabel={Number of info bits},
xmajorgrids,
xmin=5, xmax=10,
xtick style={color=black},
xtick={3,4,5,6,7,8,9,10},
xticklabels={8,16,32,64,128,256,512,1024},
y grid style={darkgray176},
ylabel={SNR [dB]},
ymajorgrids,
ymin=-9.125, ymax=15.625,
ytick style={color=black}
]
\addplot [semithick, red, dash pattern=on 5.55pt off 2.4pt, mark=square, mark size=2.5, mark repeat=2, mark options={solid,fill opacity=0}]
table {%
5.67807190511264 -8
6.67807190511264 -5.5
7.26303440583379 -3
7.67807190511264 -1.5
8 0
8.26303440583379 1
8.48542682717024 2
8.67807190511264 3.5
8.84799690655495 4.5
9 6
9.13750352374993 7.5
9.26303440583379 9
9.37851162325373 10.5
};
\addlegendentry{NPD Ideal HPA}
\addplot [semithick, blue, mark=triangle*, mark size=2.5, mark repeat=2, mark options={solid}]
table {%
5.67807190511264 -3.5
6.67807190511264 -2
7.26303440583379 -0.5
7.67807190511264 0.5
8 1.5
8.26303440583379 2.5
8.48542682717024 4
8.67807190511264 5
8.84799690655495 6
9 7.5
9.13750352374993 9
9.26303440583379 10
9.37851162325373 11.5
};
\addlegendentry{5G Polar Ideal HPA}
\addplot [semithick, darkviolet1910191, mark=triangle*, mark size=2.5, mark repeat=2, mark options={solid,rotate=180}]
table {%
5.67807190511264 -0.5
6.67807190511264 1
7.26303440583379 2.5
7.67807190511264 3.5
8 5
8.26303440583379 6
8.48542682717024 7
8.67807190511264 8.5
8.84799690655495 9.5
9 11
9.13750352374993 12.5
9.26303440583379 13
9.37851162325373 15.5

};
\addlegendentry{5G Polar IBO=0}
\addplot [semithick, green01270, dash pattern=on 5.55pt off 2.4pt, mark=triangle, mark size=2.5, mark repeat=2, mark options={solid,rotate=90,fill opacity=0}]
table {%
5.67807190511264 -5
6.67807190511264 -2
7.26303440583379 0
7.67807190511264 1.5
8 3
8.26303440583379 4.5
8.48542682717024 5.5
8.67807190511264 7
8.84799690655495 8
9 9.5
9.13750352374993 11
9.26303440583379 13
9.37851162325373 14.5
};
\addlegendentry{NPD IBO=0}
\end{axis}

\end{tikzpicture}}
\caption{SNR required to achieve BLER \(=10^{-3}\) under nonlinear HPA conditions, shown as a function of the number of information bits. Results include RAPP model with IBO = 0\,dB and an ideal linear amplifier.}
\label{fig:nonlinear_amp_plot}
\end{figure}
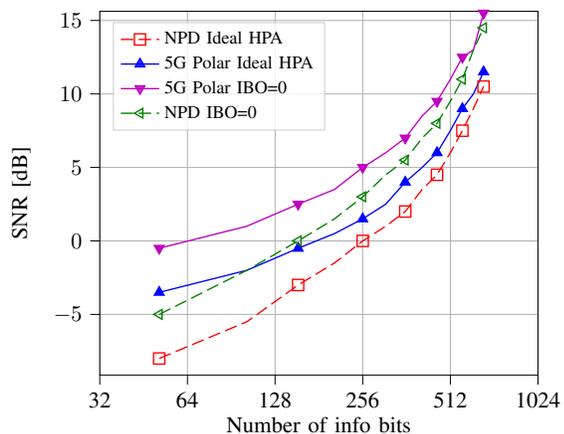

\subsection{Performance Over Nonlinear Power Amplifiers}

This experiment evaluated the performance of the NPD compared with that of the 5G polar decoder under nonlinear power amplifier (HPA) conditions. Specifically, we employed the RAPP model \cite{rapp_1} to simulate the HPA nonlinearity using a smoothness parameter \( p = 2 \). Two scenarios were considered: a nonlinear HPA with an input back-off (IBO) of 0\,dB and an ideal linear amplifier as a reference. The communication setup assumes a code length of \( 1024 \), a TDL-A channel with a 100ns delay spread, and user mobility uniformly distributed in the range 0--8m/s. BPSK modulation was used.

\textbf{Insight}: The results demonstrate that the NPD maintains its performance advantage over the 5G polar decoder even under nonlinear amplifier conditions, which are part of practical communication systems.

Figure~\ref{fig:nonlinear_amp_plot} shows the required SNR to achieve a target BLER of \( 10^{-3} \) for different numbers of information bits. As expected, both decoders exhibited performance degradation in the presence of amplifier nonlinearity. However, the NPD consistently outperformed the 5G Polar.

\subsection{Throughput Comparison}

This experiment compares the throughput of the NPD and the 5G polar decoder under identical resource element (RE) allocations per frame to ensure a fair comparison. Each frame consists of eight OFDM symbols, with the number of subcarriers determined by the total RE budget. For instance, with 256 REs per frame, the system uses 32 subcarriers. In the 5G polar decoder, the $1P_2$ and $2P_2$ configurations introduce pilot overheads of 16 and 32 REs, respectively. Throughput is computed as: $\text{Throughput} = (1 - \text{BLER}) \cdot \frac{k}{T},$ where \(k = 32\) is the number of transmitted information bits, and \(T = \frac{1}{\text{subcarrier spacing}} + T_{\text{cp}}\) denotes the total frame duration. For the 5G decoder, the CP duration is \(T_{\text{cp}} = 4.69\,\mu\text{s}\), whereas the NPD operates without a CP ($T_{cp}=0$).

\textbf{Insight}: These experiments show that the NPD significantly outperforms the 5G polar decoder in terms of throughput. This gain is attributed to the NPD’s superior BLER performance, its ability to employ longer code lengths (as the 5G decoder reserves REs for pilots), and its operation without a CP, which provides an additional throughput gain of approximately $7\%$ compared to systems using a normal CP.

These insights are drawn from Figure~\ref{fig:throughput_doppler}, which illustrates the throughput versus SNR results for both high and low Doppler conditions across various REs per frame. In the high Doppler scenario, Figure \ref{fig:thdophigh}, corresponding to user velocities between 15 and 30m/s, the 5G polar decoder employs the 2P configuration to maintain robustness under rapid channel variation. In contrast, in the low Doppler case, Figure \ref{fig:thdoplolw}, where velocities range from 0 to 8~m/s, the 1P configuration is used to improve spectral efficiency. Under both Doppler conditions, the NPD consistently achieved a higher throughput across all SNR levels and RE configurations. It begins decoding successfully at lower SNRs and continues to outperform the 5G polar decoder at higher SNRs by approximately 7\%, primarily due to the elimination of the CP, benefiting from superior decoding performance and reduced transmission overhead. 

Although the results shown correspond to a specific case with \(k=32\) information bits, similar trends were observed across a wider range of \(k\) values. This observation is also supported by the superior BLER performance of the NPD, as shown in the Figure \ref{fig:block-size-comparison}.

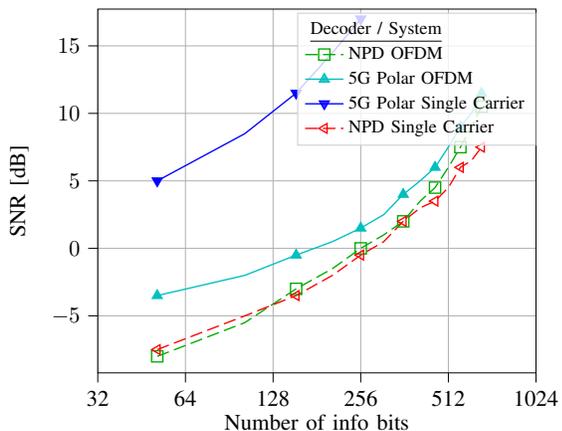
\begin{figure}[h!]
\centering
\scalebox{.85}{
\begin{tikzpicture}

\definecolor{darkgray176}{RGB}{176,176,176}
\definecolor{lightgray204}{RGB}{204,204,204}
\definecolor{darkturquoise0191191}{RGB}{0,191,191}
\definecolor{green2}{RGB}{0,170,0}
\begin{axis}[
legend cell align={left},
legend style={
  font=\footnotesize,
  fill opacity=0.8,
  draw opacity=1,
  text opacity=1,
  at={(1,0.99)},
  anchor=north east,
  draw=lightgray204
},
tick align=outside,
tick pos=left,
x grid style={darkgray176},
xlabel={Number of info bits},
xmajorgrids,
xmin=32, xmax=1024,
xtick style={color=black},
xmode=log,
log basis x=2,
xtick={8,16,32,64,128,256,512,1024},
xticklabels={8,16,32,64,128,256,512,1024},
y grid style={darkgray176},
ylabel={SNR [dB]},
ymajorgrids,
ymin=-9.225, ymax=17.725,
ytick style={color=black}
]
\addlegendimage{empty legend}
\addlegendentry{\hspace{-.6cm}\underline{Decoder / System}}
\addplot [semithick, green2, dash pattern=on 5.55pt off 2.4pt, mark=square, mark size=2.5, mark repeat=2, mark options={solid,fill opacity=0}]
table {%
51.2 -8
102.4 -5.5
153.6 -3
204.8 -1.5
256 0
307.2 1
358.4 2
409.6 3.5
460.8 4.5
512 6
563.2 7.5
614.4 9
665.6 10.5
};
\addlegendentry{NPD OFDM}
\addplot [semithick, darkturquoise0191191, mark=triangle*, mark size=2.5, mark repeat=2, mark options={solid}]
table {%
51.2 -3.5
102.4 -2
153.6 -0.5
204.8 0.5
256 1.5
307.2 2.5
358.4 4
409.6 5
460.8 6
512 7.5
563.2 9
614.4 10
665.6 11.5
};
\addlegendentry{5G Polar OFDM}
\addplot [semithick, blue, mark=triangle*, mark size=2.5, mark repeat=2, mark options={solid,rotate=180}]
table {%
51.2 5
102.4 8.5
153.6 11.5
204.8 14.5
256 17
};
\addlegendentry{5G Polar Single Carrier}
\addplot [semithick, red, dash pattern=on 5.55pt off 2.4pt, mark=triangle, mark size=2.5, mark repeat=2, mark options={solid,rotate=90,fill opacity=0}]
table {%
51.2 -7.5
102.4 -5
153.6 -3.5
204.8 -2
256 -0.5
307.2 0.5
358.4 2
409.6 3
460.8 3.5
512 4.5
563.2 6
614.4 6.5
665.6 7.5
};
\addlegendentry{NPD Single Carrier}
\end{axis}

\end{tikzpicture}}
\caption{Required SNR to achieve a target BLER of \(10^{-3}\) for single-carrier and OFDM systems over a TDL-A channel with 100\,ns delay spread and user mobility of 0–8\,m/s. All systems use BPSK modulation and a code length of \(N = 1024\).}
\label{fig:siso_rate_plot}
\end{figure}

\subsection{Comparison of Single-Carrier and OFDM Systems}

This experiment compares the performance of the NPD and 5G polar decoder in single-carrier and OFDM-based systems. The systems use BPSK modulation and a code length of $N = 1024$, with user mobility in the range of 0--8\,m/s. For the single-carrier waveform, a Sinc filter was used for pulse shaping and receive filtering, using the same bandwidth as the corresponding OFDM configuration to ensure a fair comparison.

\textbf{Insight}: The proposed NPD enables single-carrier systems to achieve decoding performance comparable to that of OFDM, even under severe multipath conditions and without increasing complexity. This is particularly impactful, given the inherent advantages of single-carrier waveforms, such as lower PAPR and simpler hardware requirements. The reduced PAPR improves the power amplifier efficiency under nonlinear conditions. These results highlight the potential of NPD-based single-carrier systems as viable alternatives to OFDM.

Figure \ref{fig:siso_rate_plot} shows the required SNR to achieve a BLER of \(10^{-3}\) as a function of the number of transmitted information bits. A significant performance gap was observed between the 5G polar decoder using a single carrier and the NPD under the same waveform. Notably, the NPD with a single carrier achieves a performance comparable to that of the NPD with OFDM.

\section{Conclusions and Future Work}
\label{sec:conc}

In this study, we presented extended NPDs tailored for practical communication systems. The proposed NPD operates directly over memory channels without relying on pilots and CP. NPDs offer a robust and reliable solution with a manageable decoding complexity. The NPD was enhanced to support higher-order modulation, rate matching for any code lengths, and robust performance across various channel conditions. Evaluations over 5G channels show that the NPD consistently outperforms the 5G polar decoder in terms of BER, BLER, and throughput. In particular, it shows significant gains in the low rate and short code length that are typical of control channels. 

Furthermore, the NPD exhibits robustness across different code rates, BPSK and QPSK modulation schemes, and channel conditions without requiring retraining. It also demonstrated resilience to nonlinear distortions caused by power amplifiers. The single carrier with NPD achieves a decoding performance comparable to that of OFDM over a memory channel. This enables single-carrier systems to operate with the same decoding complexity while offering advantages such as a lower PAPR, which improves the amplifier efficiency under nonlinear conditions.  Overall, the NPD offers a pilotless, high-performance decoding solution with a strong potential for future wireless systems.

Several promising research directions remain for future exploration. One potential avenue is the investigation of a universal reliability sequence for NPD, which could eliminate the need for scenario-specific design and enhance its practical applicability. Another important avenue of investigation is the extension of NPDs to multiple-input multiple-output (MIMO) systems, where decoding complexity significantly increases, and data-driven approaches could yield substantial performance improvements. Additionally, developing hardware-aware NPDs is a promising direction. Techniques like quantization and pruning can reduce model complexity and computation, enabling efficient deployment on resource-constrained platforms such as IoT and edge devices.

\section*{APPENDIX}
\subsection{NPDLoss}\label{sec:NPDloss}

This subsection presents the algorithm used to compute the $\mathsf{NPDLoss}$ in \(\mathcal{O}(\log N)\) steps. The method is based on the algorithm in~\cite{aharoniCodeRateOptimization2024}, with the modification that the input is the embedding $e_0$ rather than $y^N$. Consequently, the embedding function is applied outside the loss computation.

\begin{algorithm}[h!]
\footnotesize 

\caption{NPDLoss}
\begin{algorithmic}

\State \textbf{Input:} 
\Statex \hspace{1em} $e_0,x^N$ 
\Statex \hspace{1em} $F_\theta, G_\theta, H_\theta$ \Comment{\gls{npd} model}
\State \textbf{Output:} 
\Statex \hspace{1em} $\mathcal{L}(x^N, y^N; \theta)$ \Comment{computed loss}
\algrule
\State \textbf{Stage 0:}
    \State \hspace{1em}$\mathbf{v}_0 \leftarrow x^N$
    \State \hspace{1em}$\mathbf{l}_0 \leftarrow \f{\mathbf{H}_\theta}{\mathbf{e}_0}$
    \vspace{-2em}
    \State \begin{align*}
    \hspace{0.3em}\mathcal{L}_0(\mathbf{v}_0, \mathbf{e}_0; \theta)= -\frac{1}{N} \sum_{i=1}^N v_{0,i} \log \sigma(l_{0,i}) + \overline{v}_{0,i} \log \overline{\sigma(l_{0,i})} 
    \end{align*}
        \Comment{loss of stage $0$}

\State \textbf{Stage 1:}
    \State \hspace{1em}$\mathbf{v}_1 \gets [\mathbf{v}_0^o \oplus \mathbf{v}_0^e \mid \mathbf{v}_0^e]$
    \State \hspace{1em}$\mathbf{e}_1 \gets [\mathbf{F}_\theta(\mathbf{e}_0^o, \mathbf{e}_0^e) \mid \mathbf{G}_\theta(\mathbf{e}_0^o, \mathbf{e}_0^e, \mathbf{v}_0^o \oplus \mathbf{v}_0^e)]$
    \State \hspace{1em}$\mathbf{l}_1 \gets \mathbf{H}_\theta(\mathbf{e}_1)$
        \vspace{-2em}
    \State \begin{align*}
    \hspace{0.3em}\mathcal{L}_1(\mathbf{v}_1, \mathbf{e}_1; \theta)= -\frac{1}{N} \sum_{i=1}^N v_{1,i} \log \sigma(l_{1,i}) + \overline{v}_{1,i} \log \overline{\sigma(l_{1,i})} 
    \end{align*}
    \Comment{loss of stage $1$}
    
\State \textbf{Stages 2 to $\mathbf{n}$:}
\hspace{1em}\For{$j = 2$ to $n$}
    \State Compute $\mathbf{v}^{(j-1)}$ and $\mathbf{e}^{(j-1)}$ \Comment{Equation \eqref{eqn:gradient_split}}
    \State Initiate $\mathbf{v}^{(j)}=\emptyset, \mathbf{e}^{(j)}=\emptyset, \mathbf{l}^{(j)}=\emptyset$
    \For{each $\mathbf{\tilde{v}}_{j-1} \in \mathbf{v}^{(j-1)}$ and $\mathbf{\tilde{e}}_{j-1} \in \mathbf{e}^{(j-1)}$}
        \State Compute $\mathbf{\tilde{v}}_j$, $\mathbf{\tilde{e}}_j$ and $\mathbf{\tilde{l}}_j$ \Comment{Equation \eqref{eqn:gradient_v}--\eqref{eqn:gradient_l}} 
        \State $\mathbf{v}^{(j)}\gets\mathbf{v}^{(j)}\cup\mathbf{\tilde{v}}_j$, $\mathbf{e}^{(j)}\gets\mathbf{e}^{(j)}\cup\mathbf{\tilde{e}}_j$, $\mathbf{l}^{(j)}\gets\mathbf{l}^{(j)}\cup\mathbf{\tilde{l}}_j$
    \EndFor
    \State Concatenate $\mathbf{v}^{(j)},\mathbf{e}^{(j)},\mathbf{l}^{(j)}$ into $\mathbf{e}_j$, $\mathbf{v}_j$, $\mathbf{l}_j$ 
        \vspace{-2em}
    \State \begin{align*}
    \hspace{0.3em}\mathcal{L}_j(\mathbf{v}_j, \mathbf{e}_j; \theta)= -\frac{1}{N} \sum_{i=1}^N v_{j,i} \log \sigma(l_{j,i}) + \overline{v}_{j,i} \log \overline{\sigma(l_{j,i})} 
    \end{align*}
    \Comment{loss of stage $j$}
    
\EndFor

\State \Return $\mathcal{L}(e_0,x^N ; \theta) \gets \frac{1}{n+1} \sum_{j=0}^n \mathcal{L}_j(\mathbf{v}_j, \mathbf{e}_j; \theta)$

\end{algorithmic}\label{alg:npd_loss}
\end{algorithm}
\par Given $x^N$ and the embeddings $e_0$, the loss of the \gls{npd} is computed as follows. the \glspl{llr} are computed as $\mathbf{l}_0 = \f{\mathbf{H}_\theta}{\mathbf{e}_0}$. The loss of bits at stage $0$ is then computed as
\begin{align}
    \f{\cL_0}{\mathbf{v}_0,\mathbf{e}_0;\theta} &= \nn\\
    &\hspace{-2cm}-\frac{1}{N}\sum_{i=1}^N v_{0,i}\log \f{\sigma}{l_{0,i}} + \overline{v}_{0,i} \log \overline{\f{\sigma}{l_{0,i}}},
\end{align}
where $\overline{x} = 1-x$, $\sigma(x) = \frac{1}{1+e^{-x}}$ is the logistic function and $\f{P_{U_i|U^{i-1},Y^N}}{1|u^{i-1},y^N}=\f{\sigma}{\f{L_{U_i|U^{i-1},Y^N}}{u^{i-1},y^N}}$.

\par At stage $1$, the loss is computed in the following manner. First, $\mathbf{v}_1$ is computed by 
\begin{equation}\label{eqn:gradient_v}
    \mathbf{v}_1 = [\mathbf{v}_0^o \oplus \mathbf{v}_0^e \mid \mathbf{v}_0^e],
\end{equation}
where $\mathbf{v}_0^o,\mathbf{v}_0^e\in \bF_2^{\frac{N}{2}}$ contain the odd and even elements of $\mathbf{v}_0$, respectively. Next, $\mathbf{e}_1$ is computed as follows:
\begin{align}\label{eqn:gradient_e}
    \mathbf{e}_1 = [\f{\mathbf{F}_\theta}{\mathbf{e}_0^o,\mathbf{e}_0^e} \mid 
                \f{\mathbf{G}_\theta}{\mathbf{e}_0^o,\mathbf{e}_0^e, \mathbf{v}_0^o \oplus \mathbf{v}_0^e} ],
\end{align}
where $\f{\mathbf{F}_\theta}{\mathbf{e}_0^o,\mathbf{e}_0^e},\f{\mathbf{G}_\theta}{\mathbf{e}_0^o,\mathbf{e}_0^e, \mathbf{v}_0^o \oplus \mathbf{v}_0^e}\in\bR^{\frac{N}{2}\times d}$, the operator $[\cdot|\cdot]$ denotes the concatenation of two matrices along the first dimension, and $\mathbf{e}_1\in\bR^{N\times d}$. The loss of stage $1$ is then computed by first computing
\begin{equation}\label{eqn:gradient_l}
\mathbf{l}_1 = \f{\mathbf{H}_\theta}{\mathbf{e}_1}, 
\end{equation}
and then computing $\f{\cL_1}{\mathbf{v}_1,\mathbf{e}_1;\theta}$, as done in stage $0$.

\par At stages $j\in[2:n]$, the same computations as in stage $1$ are followed, but they are performed within sub-blocks independently. Given $\mathbf{e}_{j-1}$ and $\mathbf{v}_{j-1}$, we first split them into collections:
\begin{align}\label{eqn:gradient_split}
    \mathbf{v}^{(j-1)}&=\left\{\mathbf{v}_{j-1,k}\right\}_{k=0}^{2^{j-1}-1} \nn\\
    \mathbf{e}^{(j-1)}&=\left\{\mathbf{e}_{j-1,k}\right\}_{k=0}^{2^{j-1}-1},
\end{align}
where 
\begin{align}\label{eqn:gradient_split}
    \mathbf{v}_{j-1,k} &= \left\{v_{j-1,l}\right\}_{l=1+k2^{n-j+1}}^{(k+1)2^{n-j+1}} \nn\\
    \mathbf{e}_{j-1,k} &= \left\{e_{j-1,l}\right\}_{l=1+k2^{n-j+1}}^{(k+1)2^{n-j+1}}, \nn
\end{align}
with $\mathbf{v}_{j-1,k} \in \bF_2^{\frac{N}{2^{j-1}}\times 1}$ and $\mathbf{e}_{j-1,k} \in \bR^{\frac{N}{2^{j-1}}\times d}$.
For every $\mathbf{v}_{j-1,k}$ and $\mathbf{e}_{j-1,k}$, we repeat the computations in Equations \eqref{eqn:gradient_v}--\eqref{eqn:gradient_e} to produce $\mathbf{v}^\prime_{j-1,k}$ and $\mathbf{e}^\prime_{j-1,k}$, which are then concatenated into $\mathbf{e}_{j}$, $\mathbf{v}_{j}$. Next, $\mathbf{e}_{j}$, $\mathbf{v}_{j}$ are used to compute $\cL_j\left(\mathbf{v}_j,\mathbf{e}_j;\theta\right)$ and are passed to the next stage. The overall loss is computed by 
\begin{equation}\label{eqn:npd_loss_orig}
    \f{\cL}{x^n,y^n;\theta} = \frac{1}{n+1}\sum_{j=0}^n \f{\cL_j}{\mathbf{v}_j,\mathbf{e}_j;\theta},
\end{equation}
and the corresponding gradient is given by $\nabla_\theta \f{\cL}{x^n,y^n;\theta}$. The loss computation is summarized in Algorithm \ref{alg:npd_loss}.

\bibliography{ref}
\end{document}